\documentclass[aps,physrev,twocolumn,superscriptaddress,showkeys]{revtex4-2}

\usepackage{amsmath, amsthm, amssymb, amsfonts}
\usepackage{float}
\usepackage{tikzsymbols}
\usepackage[export]{adjustbox}
\usepackage{xcolor}
\usepackage{soul}
\usepackage{enumitem}
\usepackage{graphicx}
\usepackage{comment}
\graphicspath{ {PRB/figs/} }
\usepackage[T1]{fontenc}
\usepackage[english]{babel}

\newcommand{\ket}[1]{\left|#1\right>}
\newcommand{\expval}[1]{\left< #1 \right>}
\newcommand{\mel}[3]{\left< #1 \right| #2 \left| #3 \right>}

\begin{document}

\title{Quadrupolar and dipolar phases of excitons in transition-metal dichalcogenide trilayer heterostructures}

\author{Michal Zimmerman }
\email[Contact author: ]{michal.zimmerman@mail.huji.ac.il}
\affiliation{The Racah Institute of Physics, The Hebrew University of Jerusalem, Jerusalem 9190401, Israel}
\author{Daniel Podolsky }
\affiliation{Department of Physics, Technion, Haifa 3200003, Israel}
\author{Ronen Rapaport }
\affiliation{The Racah Institute of Physics, The Hebrew University of Jerusalem, Jerusalem 9190401, Israel}
\author{Snir Gazit }
\email[Contact author: ]{snir.gazit@mail.huji.ac.il}
\affiliation{The Racah Institute of Physics, The Hebrew University of Jerusalem, Jerusalem 9190401, Israel}
\affiliation{The Fritz Haber Research Center for Molecular Dynamics, The Hebrew University of Jerusalem, Jerusalem 9190401, Israel}

% \date{\today}

\begin{abstract}
Recent experiments on trilayer transition-metal dichalcogenide heterostructures have revealed the rich behavior of dipolar excitons. Motivated by these experimental observations, we investigate the collective dynamics of planar quantum dipoles whose orientations fluctuate because of charge tunneling between the outer layers. Using large-scale quantum Monte Carlo simulations, we map out the low-temperature phase diagram as a function of experimentally tunable parameters. We uncover a diverse landscape of phases driven by dipolar correlations. Under strong dipole fluctuations, a quadrupolar superfluid emerges. Suppressing charge tunneling nucleates a droplet state stabilized by the attractive interaction between antiparallel dipoles. At high exciton densities, the system gives way to a partially fragmented condensate, characterized by a finite interaction-driven occupation of the two dipolar states alongside a dominant quadrupolar superfluid component. Furthermore, exploring large effective masses that suppress zero-point motion, we find a staggered dipolar crystal at high densities. Our detailed study of the dependence of exciton energy shifts on an external electric field provides qualitative theoretical guidance for interpreting existing data, underscoring the crucial role of the antiparallel dipolar configuration and paving the way for future experimental explorations of quantum phases of trilayer excitons.
\end{abstract}

\maketitle

\section*{Introduction}
Experiments in dipolar gases of ultracold atoms~\cite{Chomaz_2023,Baranov_2012} and dipolar excitons in semiconducting heterostructures~\cite{Deng_2024,Glazov_2019} provide a unique platform for studying strongly correlated quantum phenomena. The long-range and anisotropic nature of dipolar interactions, which alternate between repulsion and attraction, enables access to a wide range of unconventional collective behavior. In excitonic systems, bilayer structures serve as the fundamental building blocks for stabilizing excitons with a finite dipole moment arising from the spatial separation of electrons and holes.

Extensive studies of spatially indirect dipolar excitons in bilayer semiconductor double quantum wells have revealed a rich landscape of collective phenomena. Under an external polarizing electric field, these systems exhibit distinct many-body correlations~\cite{Laikhtman_2009, Shilo_2013, Cohen_2016, Misra_2018}, extended-range spatial coherence~\cite{ButovCoherence2012,Dubin_2017}, and dynamical condensation into a dark ground state~\cite{Combescot_dark, Combescot_gray, Mazuz_2019, Misra_2022, Rapaport_2020, IBJ_2024}.

More recently, attention has shifted toward atomically thin transition-metal dichalcogenide (TMD) heterostructures. The structural versatility of bilayer TMD systems~\cite{Deng_2024,Holleitner_2024} has enabled the observation of Mott insulators and Wigner crystals~\cite{Mak2021EI,Mak2022EI,Watanabe2022EI,Mak2020Mott,Watanabe2020Mott,Imamoglu2020Mott,Wang2020Mott,Watanabe2020Wigner,Knap2025Diffusion}, macroscopically coherent interlayer exciton superfluidity~\cite{Cutshall_2025}, and atomic-like properties such as electrically tunable Feshbach resonances~\cite{Imamoglu2021Feshbach,Imamoglu2022Feshbach}. In bilayers, however, dipolar interactions remain purely repulsive and isotropic.

Multilayer structures overcome this limitation by exposing the full anisotropy of dipolar interactions through a competition between in-plane repulsion and out-of-plane attraction, recently observed experimentally~\cite{Hubert_2019,Rapaport_2020,Choksy_2021}. The attractive component can stabilize exotic phases, including pair- and trimer-superfluidity and paired crystals~\cite{Zimmerman_2022,Cinti_2017,Belgaonkar_2025,Rapaport_2020,Hubert_2019,Fogler_2021}. The simplest non-trivial extension of bilayers is the trilayer TMD structure, which has recently attracted considerable experimental and theoretical interest~\cite{Slobodkin_2020,Deilmann_2024,Zhang_2024,Yu_2023,Li_2023,Bai_2023,Lian_2023,Xie_2023,Xie_2024}.

\begin{figure*}[t!]
    \centering
    \includegraphics[width=0.94\textwidth]{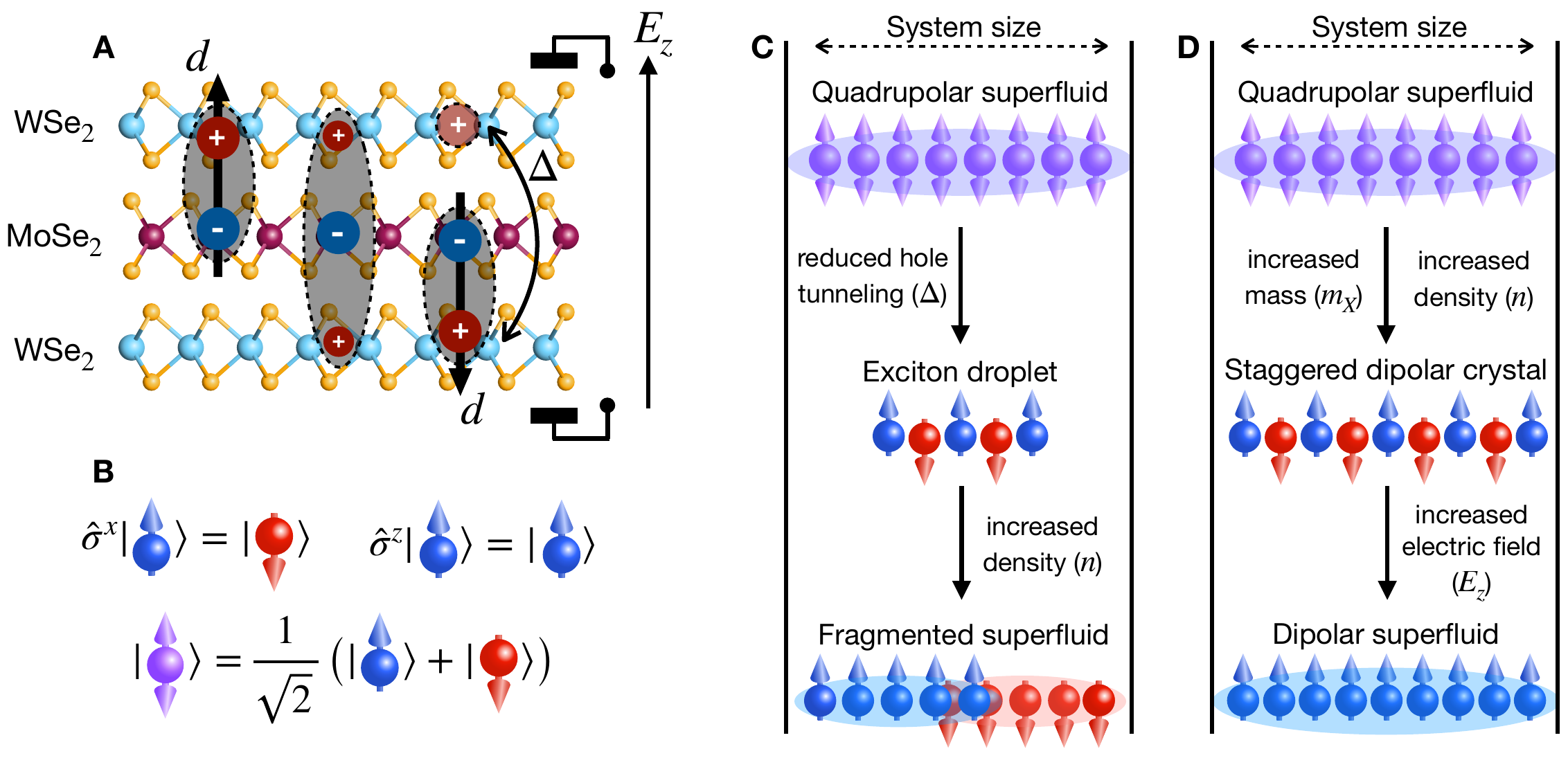}
  \caption{\textbf{Trilayer system, model and phases.} (\textbf{A}) Side-view schematic of a representative WSe$_2$/MoSe$_2$/WSe$_2$ trilayer heterostructure. The two dipolar exciton eigenstates carry opposite dipole moments, pointing up or down, based on the hole location. The quadrupolar exciton state is an equal superposition of the hole states across the outer layers. The energy splitting between the symmetric quadrupolar ground state and the excited dipolar states is given by the hole tunneling rate, $\Delta$. An experimental bias is introduced via an external out-of-plane electric field $E_z$ applied by gating the outer layers. (\textbf{B}) In the dipolar basis representation, $\hat{\sigma}^{x}$ generates dipole flips corresponding to hole tunneling events, and $\hat{\sigma}^{z}$ describes the coupling of dipoles to the external electric field. The quadrupolar state is the low-energy eigenstate of $\hat{\sigma}^{x}$, and is thus given by the symmetric superposition of the two dipolar states. (\textbf{C}) For a realistic exciton mass, $m_X=m_0$, a quadrupolar superfluid phase is realized for the currently studied experimental systems. At reduced tunneling rates, an exciton droplet with significant antiparallel dipolar correlations emerges in the dilute-density limit. 
  At higher densities, the droplet melts into a superfluid showing signatures consistent with partial fragmentation, alongside a predominantly occupied quadrupolar state. To illustrate this mechanism, the panel shows the extreme case of complete fragmentation. Our numerical data, however, only indicate a tendency toward a partial occupation of the dipolar states in tandem with the quadrupolar superfluid.
  (\textbf{D}) At larger exciton masses, $m_X\gtrsim 4m_0$, and densities, a staggered dipolar crystal appears.  The crystal melts upon application of an external out-of-plane electric field because of dipole moment polarization. The resulting phase is a dipolar superfluid.}
  \label{fig:trilayer_overview}
\end{figure*}

In trilayer TMDs, electrons localize in the central layer and holes in the outer layers (or vice versa), giving rise to two species of dipolar excitons with opposite dipole moments (see Fig.~\ref{fig:trilayer_overview}A). The inter-species interaction switches from short-range repulsion to long-range attraction. Moreover, charge tunneling between the outer layers introduces quantum fluctuations of the dipole moment. At high tunneling rates, the effective dipole moment vanishes, producing a quantum quadrupolar exciton state predicted in Ref.~\cite{Slobodkin_2020} and observed experimentally shortly thereafter~\cite{Yu_2023,Li_2023,Bai_2023,Lian_2023,Xie_2023,Xie_2024,Meng_2025}. Its hallmark is a parabolic dependence of the exciton energy shifts on an externally applied out-of-plane electric field.

At high exciton densities, where interactions dominate, experiments reveal distinct deviations from this quadrupolar behavior. The exciton energy shifts exhibit trends ranging from a linear redshift~\cite{Li_2023} to a flat response~\cite{Xie_2024}, as well as a blueshift~\cite{Yu_2023,Meng_2025}. These features have been attributed to a theoretically predicted staggered dipolar state~\cite{Slobodkin_2020}.

Understanding these observations requires a detailed theory of how quadrupolar excitons evolve into dipolar excitons across density regimes. In the limits of high density or low tunneling rates, attractive antiparallel dipolar interactions are expected to dominate. Indeed, a transition from a quadrupolar lattice to staggered droplet and crystal states has been predicted~\cite{Slobodkin_2020}. However, this prediction relies on an infinite-mass approximation that neglects finite exciton mass, quantum fluctuations~\cite{Astrakharchik_2007,Buchler_2007}, and bosonic statistics. Thus, the quantum phase diagram of trilayer excitons at high density remains unresolved.

Here, we use numerically exact, large-scale quantum Monte Carlo (QMC) simulations to determine the low-temperature phase diagram of trilayer excitons as a function of exciton density and out-of-plane electric field.  We observe several different phases, depending on the parameter regime, as summarized in Fig.~\ref{fig:trilayer_overview}.  

In the dilute, fast-tunneling regime, we observe a quadrupolar condensate.  We show that, as densities increase, antiparallel dipolar correlations emerge, directly influencing experimentally measured energy shifts.  For reduced tunneling rates, we identify a self-bound exciton droplet stabilized by antiparallel dipolar attraction. This droplet melts into a superfluid at higher densities. Notably, the resulting condensate is not purely quadrupolar but also contains a finite dipolar component, suggesting a tendency toward a partially fragmented superfluid ~\cite {Baym_2006}. Finally, at sufficiently large exciton mass, we observe a staggered dipolar crystal~\cite{Slobodkin_2020}. An applied electric field melts this crystal into a polarized superfluid, leaving a distinct fingerprint in the exciton energy shifts. We discuss the implications of our results for ongoing and near-term experiments on trilayer dipolar excitons.

\section*{Model and methods}
To study the trilayer system, illustrated in Fig.~\ref{fig:trilayer_overview}A, we consider an effective bosonic model of the excitons, following Ref.~\cite{Slobodkin_2020}. A distinctive feature of our construction is the introduction of an Ising degree of freedom assigned to each dipole moment, $\sigma^z=\pm1$, which labels the dipole orientation $\ket{\uparrow}/ \ket{\downarrow}$. The Hamiltonian reads

\begin{equation} \label{eqn:H}
    \begin{split}
        H = &-\frac{\hbar^2}{2m_X} \sum_{i=1}^N \nabla_i^2 + \sum_{i<j} U^d_{\sigma^z_i,\sigma^z_j}(|\mathbf{r}_i - \mathbf{r}_j|) \\ &- \Delta \sum_{i=1}^N \hat{\sigma}_i^x- edE_z \sum_{i=1}^N \hat{\sigma}_i^z \,.
    \end{split}
\end{equation}
Here, $i$ labels the excitons. The first term accounts for the kinetic energy of excitons with an effective mass $m_X$. The second term describes the dipolar interaction, which explicitly depends on the relative dipole orientations,
\begin{equation} \label{eqn:U_d_sigma}
    \begin{split}
        &U^d_{\sigma^z_i,\sigma^z_j}(r_{ij}) = U_{d,d}(r_{ij}) \delta_{\sigma^z_i,\sigma^z_j} + U_{d,-d}(r_{ij}) \delta_{\sigma^z_i,-\sigma^z_j}\\
        &U_{d,d}(r) = \frac{e^2}{\kappa}\left(\frac{2}{r} - \frac{2}{\sqrt{r^2 + d^2}} \right)\\
        &U_{d,-d}(r) = \frac{e^2}{\kappa}\left(\frac{1}{r} + \frac{1}{\sqrt{r^2 + (2d)^2}} - \frac{2}{\sqrt{r^2 + d^2}} \right),
    \end{split}
\end{equation}
where $r_{ij}$ is the in-plane distance between the excitons $i$ and $j$, $d$ is the dipole length, and $\kappa$ is the dielectric constant. The interaction between excitons with parallel dipole moments ($U_{d,d}$) is purely repulsive, whereas antiparallel dipoles ($U_{d,-d}$) exhibit repulsion at short separations and attraction at long distances. 

The single-exciton dynamics associated with the dipole moment orientation is captured by the last two terms of Eq.~\ref{eqn:H} and is summarized in Figs.~\ref{fig:trilayer_overview}(A) and \ref{fig:trilayer_overview}(B). The first term induces dipole orientation flipping events, since $\hat{\sigma}^x \ket{\uparrow/\downarrow}=\ket{\downarrow/\uparrow}$ when acting on the single-exciton dipolar basis. The coupling strength $\Delta$ equals the energy splitting between the quadrupolar and dipolar states~\cite{Slobodkin_2020}, thus quantifying the tunneling rate of charges between the outer layers. Tunneling between adjacent layers can be neglected, as the specific band alignment of the trilayer imposes a large energy penalty for this process \cite{Kang_2013}. The second term accounts for an external voltage applied between the outer layers, which induces an out-of-plane electric field $E_z$ along the $z$ axis. This results in a dipolar energy gain for dipoles oriented along $E_z$, such that  $edE_z \hat{\sigma}^z \ket{\uparrow/\downarrow}=\pm edE_z \ket{\uparrow/\downarrow}$.

From a symmetry perspective, for $E_z=0$, the Hamiltonian is invariant under a $\mathbb{Z}_2$ symmetry corresponding to a global dipolar orientation flip  $\sigma^z\rightarrow-\sigma^z$. Microscopically, this symmetry originates from the mirror symmetry along the central layer of the sample, and is explicitly broken for any finite $E_z$.

The characteristic energy and length scales of the model are set by the dipole length $d$ and the Coulomb coupling $\varepsilon=\frac{e^2}{\kappa d}$. Thus, the competition between dipole moment fluctuations and dipolar interactions is set by the tunneling rate $\Delta$ (in units of $\varepsilon$), and the exciton density $n$ (in units of $d^{-2}$). The relative importance of the exciton kinetic energy is governed by $E_K=\frac{\hbar^2 }{2m_X d^2 }$, expressed in units of $\varepsilon$. 
Since the dipole length and dielectric constant enter the Hamiltonian solely through these characteristic scales, they dictate the exact quantitative values of energy and density. However, varying these material parameters does not qualitatively affect the underlying physics of the model.

To determine the low-temperature phase diagram of the many-body Hamiltonian in Eq.~\ref{eqn:H}, we apply the path integral Monte Carlo approach~\cite{Ceperley_1995}, utilizing the highly efficient worm algorithm~\cite{Boninsegni_2006}. Additional technical details pertinent to our specific problem, as well as observable definitions, are detailed in Appendix~\ref{app:sim}. 

\section*{Results}
We initiate our study by considering typical experimental parameters for trilayer systems: exciton dipole length $d=0.6$~nm, exciton mass $m_X=m_0$ (with $m_0$ being the free electron mass) and dielectric constant $\kappa=2.26$~\cite{Xie_2023,Bai_2023,Yu_2023,Li_2023,Lian_2023,Slobodkin_2020,Goryca_2019,Bishwajit_2017,MacDonald_2018,Heinz_2022}, corresponding to the Coulomb energy scale $\varepsilon=1063$~meV. 

Our analysis primarily focuses on exciton densities below the critical Mott threshold, above which excitons dissociate into an unbound electron-hole plasma. Considering a typical Bohr radius of $1$~nm in TMDs~\cite{Deng_2024,Lambrecht2012biMoSe2,Berkelbach2013mono}, we infer an order of magnitude estimate for the Mott density $n_{\text{Mott}}\lesssim 10^{14}$~cm$^{-2}$. See also experimental estimates of $n_{\text{Mott}}$ in bilayer geometries~\cite{Wang_2019}. In the following, we consider, in certain cases, densities greater than $n_{\text{Mott}}$ to explore the different limits of our model, which exceed current experimental parameters.

\begin{figure*}[t!]
    \centering
    \includegraphics[width=0.95\textwidth]{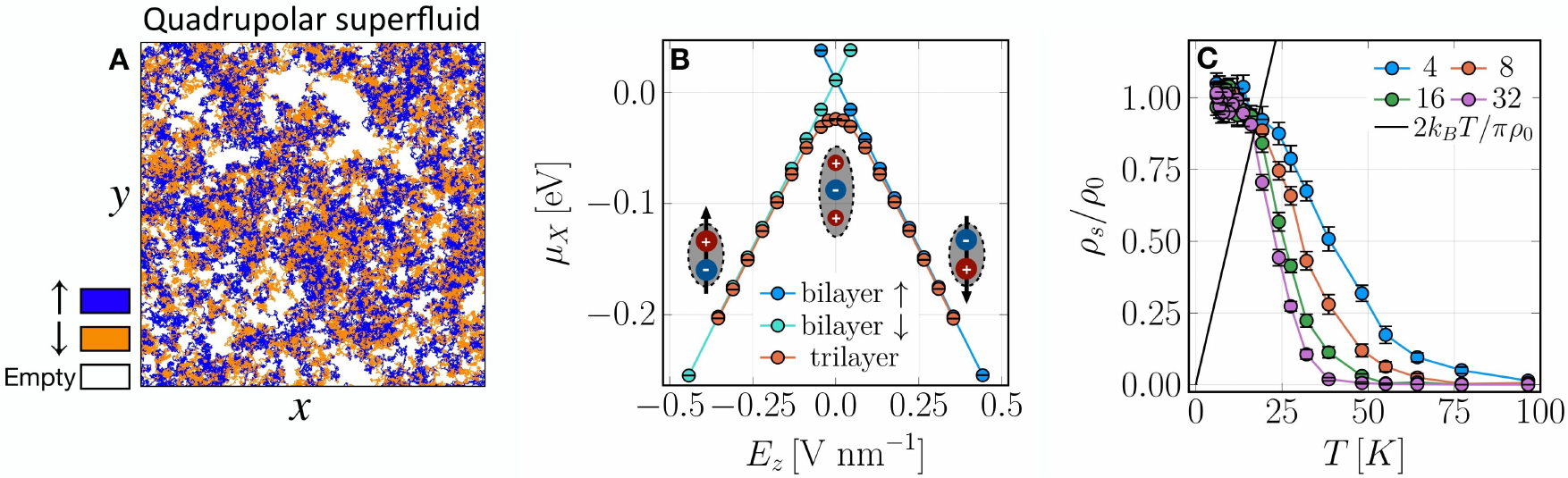}

  \caption{\textbf{Quadrupolar superfluid regime.} (\textbf{A}) A typical QMC world-line snapshot illustrating the quadrupolar superfluid state at temperature $T=6$~K. The black frame denotes the simulation box edges, to which periodic boundary conditions apply. Blue (orange) indicates the spatial positions of $\ket{\uparrow}$($\ket{\downarrow}$) excitons with projected imaginary time evolution, while white regions indicate empty regions within the box. (\textbf{B}) The exciton chemical potential as a function of the out-of-plane electric field in trilayer (red) and bilayer (blue and turquoise) geometries, shown at $T=24$~K, which probes ground-state properties (see Appendix~\ref{app:QMC}). (\textbf{C}) A finite-size analysis of the normalized superfluid stiffness in the quadrupolar phase, showing a BKT transition at the critical temperature $T_{\text{BKT}} \approx 17$~K. Data points represent different system sizes, and the black line shows the linear function $y=2 k_B T/\pi\rho_0$, divided by the normalizing factor $\rho_0$. In all panels, the exciton density is fixed at $n=1.4\cdot10^{12}$~cm$^{-2}$, with (\textbf{A}) and (\textbf{B}) computed for $N=64$ excitons. 
  }
  \label{fig:QP_SF}
\end{figure*}

\subsection*{Quadrupolar superfluid}
With the above parameters, we begin our analysis by considering a typical exciton density $n=1.4\cdot 10^{12}$~cm$^{-2}$ and a dipolar-quadrupolar energy gap $\Delta=26$~meV~\cite{Slobodkin_2020}.
To determine the resulting excitonic state, we compute the exciton chemical potential $\mu_X$~\cite{Del_Maestro_2014,Zimmerman_2022}, which is directly related to the spectral shifts of photon energies emitted in electron-hole recombination events \footnote{Assuming the emitted photon energy equals the energy difference between $N+1$ and $N$ excitons~\cite{Zimmerman_2022,Rapaport_2020,Fogler_2021}, ignoring a small fraction of inelastic losses}. Thus, $\mu_X$ serves as the main experimental probe for the excitonic state. In Fig.~\ref{fig:QP_SF}B, we plot $\mu_X$ as a function of the out-of-plane electric field $E_z$. Intriguingly, for sufficiently weak $E_z$, we observe a quadratic form, matching the anticipated energy dependence of a quadrupolar exciton subjected to an external electric field. For stronger $E_z$, the dipoles polarize and $\mu_X$ displays a linearly decreasing trend, as expected from excitons with a finite dipole moment aligned along $E_z$. This polarization effect is also clear from the convergence to the bilayer behavior, corresponding to oriented dipoles, in the large $E_z$ limit. 

The symmetric and quadratic form of $\mu_X (E_z)$ is a clear signature of a quadrupolar excitonic state and qualitatively agrees with the experimental results listed above. Importantly, the quadrupolar state is of a quantum nature, as it exhibits quantum fluctuations of dipolar orientation (Fig.~\ref{fig:trilayer_overview}B). Quantitatively, dipolar fluctuations are captured by an exponential decay of dipole-moment correlations, as we show in Appendix~\ref{app:QMC}. The decay rate is in agreement with the microscopic gap $\Delta$ for a single-exciton flip appearing in Eq.~\ref{eqn:H}.

In the low-density quadrupolar regime, excitons interact via an effective quadrupolar potential, $V(r)\sim r^{-5}$. This weakly interacting Bose gas is expected to condense into a superfluid at sufficiently low temperatures, below the Berezinskii–Kosterlitz–Thouless (BKT) transition~\cite{Astrakharchik_2021}. To explore the anticipated condensate, we examine the superfluid stiffness, $\rho_{s}$,~\cite{Ceperley_1995} as a function of temperature for an increasing range of system sizes, as shown in Fig.~\ref{fig:QP_SF}C. We find that $\rho_s$ vanishes at high temperatures and rises below the critical BKT temperature $T<T_{\text{BKT}}$, tending to its zero-temperature value, $\rho_{0}=\hbar^2n/m_X$. To extract $T_{\text{BKT}}$, we employ Nelson's jump criterion~\cite{Nelson_1977}, identifying the crossing point of the $\rho_{s}$ curves with the linear function $y=\frac{2}{\pi} k_B T $. This analysis yields $T_{\text{BKT}} \approx 17$~K, which allows experimental access to this Bose-degenerate state~\cite{Cutshall_2025}. The resulting quadrupolar superfluid is depicted by a typical snapshot of the QMC world-line path configuration in Fig.~\ref{fig:QP_SF}A. 

\begin{figure*}[t!]
    \centering
    \includegraphics[width=0.95\textwidth]{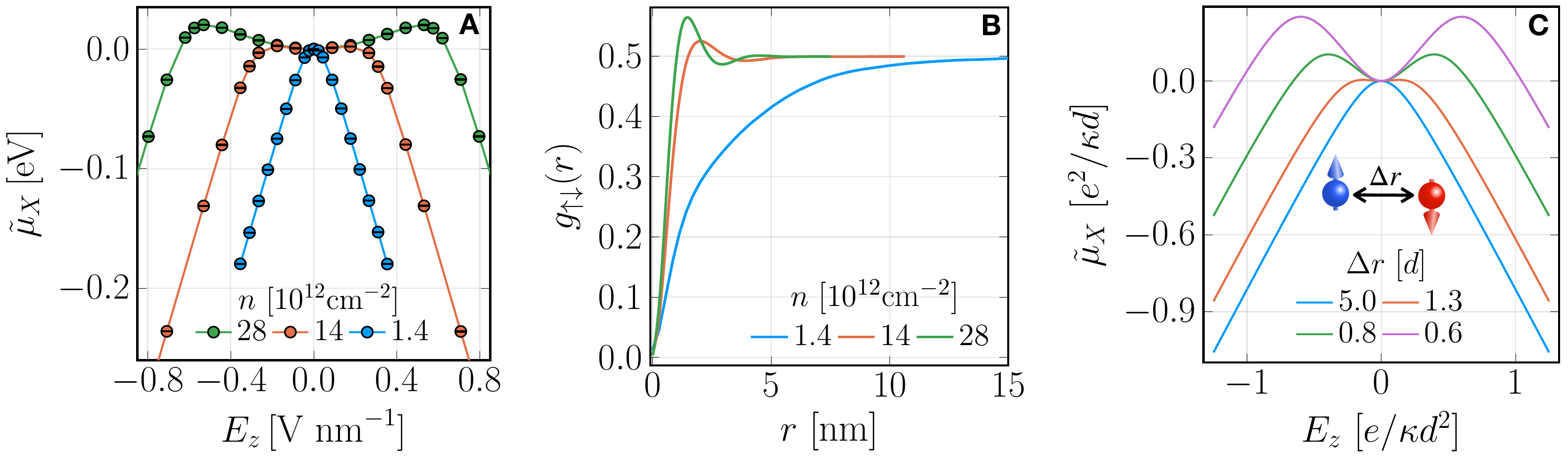}

  \caption{\textbf{Build up of antiparallel correlation as a function of density.} (\textbf{A}) Exciton chemical potential, offset by its unbiased value, as a function of the out-of-plane electric field, shown for $\Delta=26$~meV and several exciton densities spanning the low- and high-density regimes. Calculations were performed for $N=64$ excitons and in the low-temperature limit, see Appendix~\ref{app:QMC}. (\textbf{B}) The spatial correlation function between opposite dipoles, computed at $T=6$~K and for the same exciton densities and system size as in (\textbf{A}). (\textbf{C}) $\tilde{\mu}_X(E_z)$ obtained for a static two-body model of opposite dipoles, using $\Delta=0.2\varepsilon$. Each color corresponds to a fixed in-plane interexciton distance $\Delta r$.}
  \label{fig:QP_n_evol}
\end{figure*}

Thus far, we have considered the dilute-density regime, where dipole moment fluctuations stabilize a quadrupolar excitonic state. By contrast, at higher densities, excitons can benefit from the attractive component of the dipolar interaction by forming antiparallel dipolar configurations that compete with the quadrupolar state. To study this effect, we examine the dependence of the exciton chemical potential on the out-of-plane electric field, shifted by its unbiased value, $\tilde{\mu}_X(E_z) = \mu_X(E_z) - \mu_X(0)$, for an increasing range of exciton densities, as shown in Fig.~\ref{fig:QP_n_evol}A. We observe an evolution from a quadrupolar parabolic redshift in the low-density limit to a blueshift for weak $E_z$ at high densities. 

To trace the origin of this behavior, we analyze the dipole-orientation-resolved spatial correlation function of excitons, defined by 
\begin{equation} \label{eqn:gr_def}
    g_{\sigma \sigma'}(r)= \frac{A}{N^2}\expval{\frac{1}{2\pi r}\sum_{i_\sigma,j_{\sigma'}}\delta\left(r-\left|\hat{\mathbf{r}}_{i_\sigma} - \hat{\mathbf{r}}_{j_{\sigma'}}\right|\right)} \, ,
\end{equation}
where $A$ is the area. 

Figure~\ref{fig:QP_n_evol}B shows $g_{\uparrow\downarrow}(r)$ for a series of increasing densities at $E_z=0$. In the dilute limit, antiparallel dipolar correlations are absent, but they progressively emerge with increasing density, as evidenced by the development of a correlation peak in $g_{\uparrow\downarrow}(r)$. Notably, the onset of these correlations coincides with the blueshift trend observed in $\mu_X(E_z)$.

This observation suggests the following physical interpretation for the redshift and blueshift behaviors: at zero field and high densities, to minimize the energy cost of adding an exciton, it is favorable to form an antiparallel dipolar arrangement that benefits from dipolar attraction. However, as $E_z$ increases and the dipoles gradually polarize, the interexciton interaction becomes purely repulsive. This, in turn, leads to an increase in the exciton chemical potential and to the observed blueshift at finite electric field values. For a sufficiently strong electric field, the dipoles fully polarize, leading to the characteristic linear redshift expected for dipoles coupled to an external electric field. We emphasize that the appearance of antiparallel correlations away from the quadrupolar state is a crossover, since there is no sharp symmetry distinction between the two states.

To develop qualitative physical intuition for the above scenario, we examine a simple microscopic model consisting of a pair of dipolar excitons separated by a fixed relative distance $\Delta r$ and interacting through the dipolar potential given in Eq.~\ref{eqn:U_d_sigma}. In this framework, $\Delta r$ represents the mean inter-exciton distance and thus controls the interaction strength. We also allow for dipole moment orientation flips and include the coupling of dipoles to an external out-of-plane electric field. To compute the exciton energy shifts, we calculate the ground-state energy difference between the two- and single-body problems, which we solve by exact diagonalization (see Appendix~\ref{app:ED} for more details). 

In Fig.~\ref{fig:QP_n_evol}C we present $\tilde{\mu}_X(E_z)$ for several relative distances $\Delta r$. Our simple model indeed qualitatively reproduces the main trends of the many-body calculation. In particular, we observe a quadrupolar behavior at large separations, which gives way to a low-field blueshift trend at small distances. Furthermore, in Appendix~\ref{app:ED}, we explicitly demonstrate that the above behavior arises directly from the interplay between dipolar fluctuations and dipolar interactions. While this two-body picture cannot fully capture the many-body dynamics of the system, its qualitative agreement with the QMC results provides an intuitive understanding of the mechanisms driving the crossover.

\subsection*{Exciton droplet and partially fragmented  superfluid}
The parameter regimes studied above correspond to relatively large $\Delta$ values, which are associated with rapid tunneling rates. An alternative route to expose dipolar interactions is to suppress charge tunneling between the outer layers.

Following this reasoning, we consider a smaller single-particle gap of $\Delta=1$~meV. Although this value is lower than those estimated for trilayer TMD structures~\cite{Slobodkin_2020}, a reduced charge tunneling rate between the outer layers can be achieved experimentally by introducing intermediate buffer layers or a twist angle between adjacent layers~\cite{Fogler_2018,WatanabeTwist2021}.

\begin{figure*}[t!]
    \centering
    \includegraphics[width=0.85\textwidth]{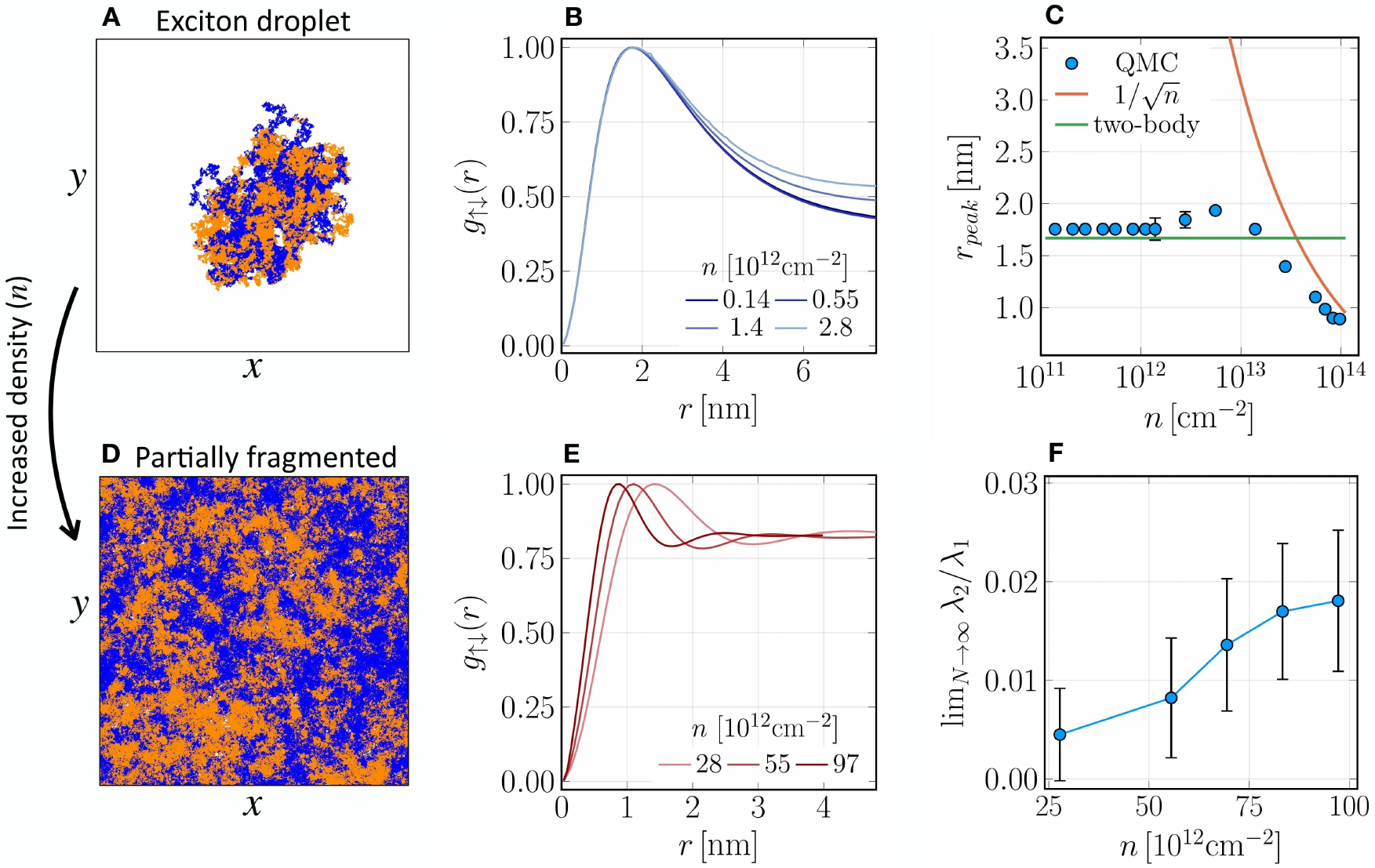}

  \caption{\textbf{Droplet and partially fragmented superfluid in the low $\Delta$ limit.} Panels (\textbf{A}) and (\textbf{D}) show typical QMC world-line snapshots of a droplet and a partially fragmented superfluid, obtained at exciton densities $n=\{0.14,28\}\cdot10^{12}$~cm$^{-2}$, temperatures $T=\{1.5,24\}$~K, and system sizes $N=\{32,256\}$, respectively. Panels (\textbf{B}) and \textbf{(E)} display spatial pair correlation functions $g_{\uparrow\downarrow}(r)$ between excitons with opposite dipole moments, for the droplet and partially fragmented superfluid states, respectively, at several densities within each phase. The curves are normalized to their maxima and computed for $N=64$ and $T=1.5$~K. (\textbf{C}) The location of the main peak of $g_{\uparrow\downarrow}(r)$ as a function of the density (blue dots), on a semilogarithmic scale, compared with the mean inter-exciton distance $1/\sqrt{n}$ (red line). The green line indicates the interparticle separation obtained from the exact two-body solution for opposite dipoles. (\textbf{F}) The thermodynamic limit of the eigenvalue ratio $\lambda_
  2/\lambda_1$, corresponding to the dipole-orientation-resolved superfluid fraction at high densities within the partially fragmented superfluid phase, shown for $T=24$~K. 
  }
  \label{fig:DP_phases}
\end{figure*}

We begin our analysis by exploring the low-density regime, setting $n=1.4\cdot10^{11}$~cm$^{-2}$. To gain a qualitative understanding of the resulting exciton state, Fig.~\ref{fig:DP_phases}A presents a typical QMC world-line snapshot of the spatial configuration. Remarkably, the system forms a self-bound exciton droplet state, characterized by pronounced spatial bunching of the excitons.

To provide a more quantitative measure of the droplet phase, we compute the antiparallel component of the spatial correlation function, $g_{\uparrow\downarrow}(r)$, defined above. In Fig.~\ref{fig:DP_phases}B, we track the evolution of $g_{\uparrow\downarrow}(r)$ as a function of exciton density, normalizing each curve by its maximum value to facilitate comparison. For densities up to $n=1.4\cdot10^{12}$~cm$^{-2}$, we observe that the primary peak in $g_{\uparrow\downarrow}(r)$ is {\it independent} of the density, suggesting an incompressible, phase-separated droplet state. Moreover, the droplet exhibits strong antiparallel correlations, as evidenced by the suppression of parallel correlations at short distances (see Appendix~\ref{app:QMC}).

At higher densities, however, the droplet is expected to destabilize because of the growing dipolar repulsion at short distances. Indeed, in the typical world-line snapshot presented in Fig.~\ref{fig:DP_phases}D for $n=2.8\cdot10^{13}$~cm$^{-2}$, we observe a qualitative change in the excitons' spatial distribution, where excitons are no longer clustered in space but instead form a liquid state. Quantitatively, the transition manifests as a continuous narrowing of $g_{\uparrow\downarrow}(r)$ with increasing density, as seen in Fig.~\ref{fig:DP_phases}E, reflecting spatial configurations featuring shorter interparticle separations. 

To further study the droplet's properties and breakdown, we analyze in Fig.~\ref{fig:DP_phases}C the evolution of the typical interparticle separation, extracted from the position of the primary peak in $g_{\uparrow\downarrow}(r)$, as a function of the exciton density. In the dilute limit, associated with the droplet state, the interexciton distance is density independent and equals $r\approx 1.75$~nm. To gain physical insight, we compare this value with that obtained from the two-body limit of Eq.~\ref{eqn:H}. For vanishing $\Delta$, the two-body ground state is an antiparallel dipolar bound state with an interparticle separation of $r=1.67$~nm (see details in Appendix~\ref{app:ED}). This value should be contrasted with the much larger naive interparticle separation determined by the lowest considered density,  $r_{\text{den}}\approx 27$~nm. The correspondence between the QMC calculation and the two-body problem underpins the role of antiparallel dipolar attraction in stabilizing the droplet. As the density increases, the interexciton separation deviates from the low-density behavior and regains the expected $1/\sqrt{n}$ scaling of a liquid. This behavior signals the breakdown of the droplet state. 

A closer inspection of the high-density liquid phase reveals an exciton condensate at sufficiently low temperatures, as evident from the finite superfluid stiffness measured in our QMC simulations (see Appendix~\ref{app:QMC}). The condensate is further decorated by the dipole moment orientation degree of freedom. To probe its internal structure, we compute the dipole-orientation-resolved superfluid fraction, 
\begin{equation}
    n_{\sigma\sigma'}(k=0)=\expval{\psi_{\sigma}(k=0)\psi^{\dagger}_{\sigma'}(k=0)} \, ,
\end{equation}
where $\psi^{\dagger}_{\sigma}(k=0)$ ($\psi_{\sigma}(k=0)$) creates (annihilates) a boson at momentum $k=0$ with dipole orientation $\sigma^{x}$. The quantity $n_{\sigma\sigma'}(k=0)$ acts as a matrix order parameter for the single-particle condensate. 

The two eigenvalues of $n_{\sigma\sigma'}(k=0)$, denoted by $\lambda_1$ and $\lambda_2$ (where $\lambda_1>\lambda_2$), carry the following physical interpretation. The limit $\lambda_2/ \lambda_1 \to 0$ corresponds to a pure condensate, where all excitons populate the same eigenstate in the internal space. In the context of our model, this scenario occurs in the quadrupolar superfluid state obtained above. Explicitly, the quantum wave function associated with the dipole moment orientation degree of freedom is the symmetric combination $\ket{\psi_{\text{quad}}}=\frac{1}{\sqrt{2}}\left(\ket{\uparrow}+\ket{\downarrow}\right)$, as dictated by minimization of the dipole-flip term. 

Away from this limit, when $\lambda_2 / \lambda_1$ remains finite, dipolar interactions compete with the quadrupolar configuration, favoring partial occupation of dipolar states. Consequently, the high-density superfluid phase may acquire a more complex internal structure, analogous to fragmented superfluidity in multispecies condensates~\cite{Penrose_1956,Baym_2006}. In the present context, the different species correspond to distinct dipole orientations. 

With this in mind, in Fig.~\ref{fig:DP_phases}F we show the eigenvalue ratio $\lambda_2/\lambda_1$ extrapolated to the thermodynamic limit ($N\to\infty$), as detailed in Appendix~\ref{app:QMC}. Considering several exciton densities within the superfluid region, we find that $\lambda_2/\lambda_1$ attains a small but finite value that increases with density. This result suggests partial fragmentation, deviating from the pure quadrupolar limit ($\lambda_2/\lambda_1=0$), driven by antiparallel dipolar correlations at high exciton densities.

While our data are consistent with partial fragmentation, the values obtained for the ratio $\lambda_2/\lambda_1$ are small ($\sim 0.02$) and thus suggest that alternative scenarios, such as a finite-size crossover, where the apparent fragmentation ultimately vanishes as $N \to \infty$, remain plausible. Therefore, we interpret this result as a tendency toward partial fragmentation, leaving the conclusive classification of the thermodynamic limit to future studies.

Finally, to ensure that thermal fluctuations have a negligible effect on the above-reported quantum phases, our simulation temperatures were verified to represent the system's ground-state limit, as detailed in Appendix~\ref{app:QMC}.

\subsection*{Staggered dipolar crystal}

Within the above parameter regime, we did not observe the staggered crystalline phase predicted in~\cite{Slobodkin_2020}, even when considering relatively high densities (Appendix~\ref{app:QMC}). We attribute this result to the large zero-point motion of the excitons, which destabilizes the crystal phase~\cite{Nelson_2014}.

\begin{figure*}
    \centering
    \includegraphics[width=1.\textwidth]{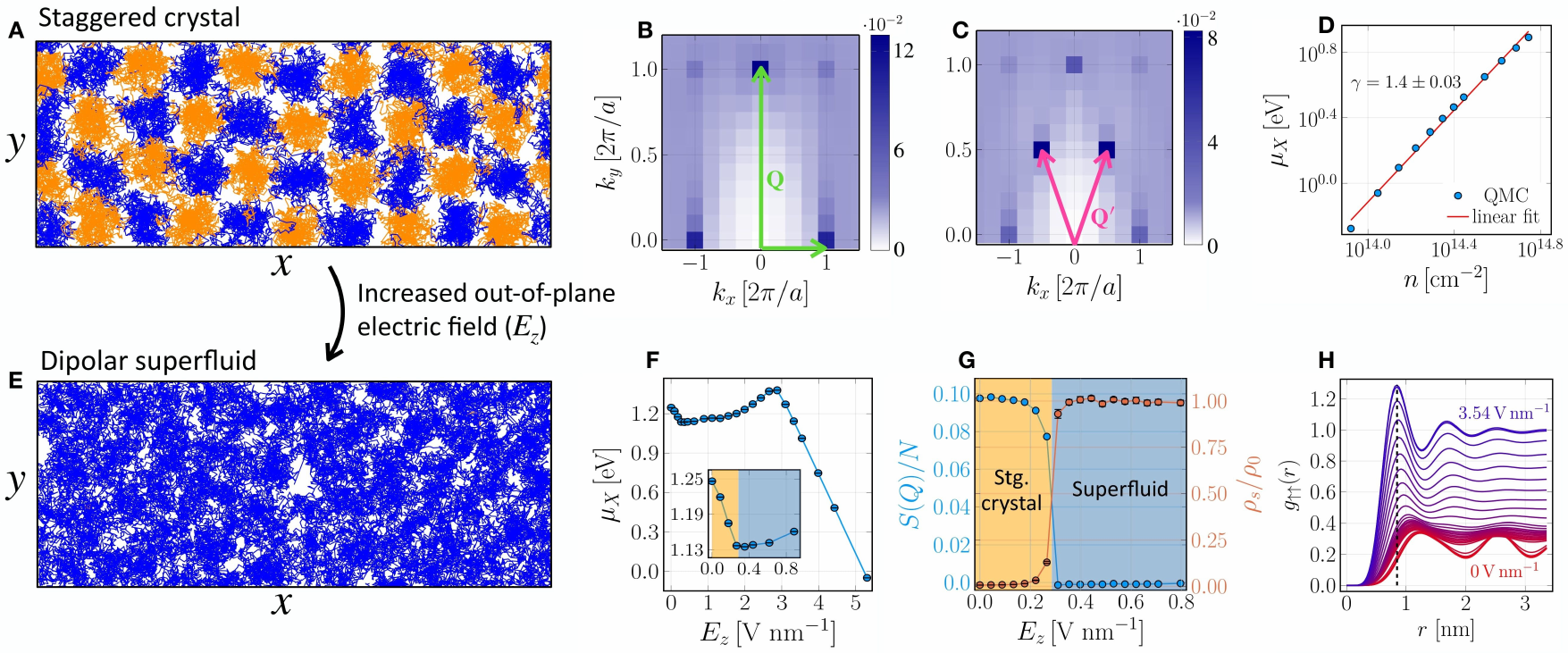}

  \caption{\textbf{The staggered crystal and its electric-field-induced melting.} (\textbf{A}) World-line snapshot of the crystal at exciton density $n=1.4\cdot10^{14}$~cm$^{-2}$. The staggered pattern is identified from the momentum-space maps of (\textbf{B}) $S(\mathbf{k})$ and (\textbf{C}) $S_{\sigma}(\mathbf{k})$, which peak at distinct Bragg vectors, $\mathbf{Q}$ and $\mathbf{Q}'$, corresponding to two square lattices with a relative $\pi/4$ rotation. (\textbf{D}) Exciton chemical potential within the crystalline phase (see Appendix~\ref{app:QMC}), plotted as a function of density on a logarithmic scale. The red line shows a linear fit indicating a power law scaling, $\mu_X\propto n^{\gamma}$, with $\gamma=1.40\pm0.03$. (\textbf{E}) World-line snapshot showing the melting of the staggered crystal in (\textbf{A}) into a dipolar superfluid under a strong electric bias $E_z=2$~V\,nm$^{-1}$. (\textbf{F}) Exciton chemical potential and (\textbf{G}) superfluid stiffness (normalized by $\rho_0$, blue) together with the Bragg peak amplitude $S(\mathbf{Q})$ (red), plotted as a function of $E_z$. The inset of (\textbf{F}) highlights the onset of melting at low $E_z$. Orange and blue backgrounds in (\textbf{F}) and (\textbf{G}) indicate the staggered crystal and superfluid regions, respectively. (\textbf{H}) Spatial correlation function between parallel dipoles aligned with $E_z$, shown for increasing electric field values from $E_z=0$ (red) to $E_z=3.54$~V\,nm$^{-1}$ (blue). The black dashed line marks the mean inter-exciton distance $a=1/\sqrt{n}$. Panels (\textbf{A})-(\textbf{F}), as well as the $\rho_s$ curve in (\textbf{G}), were computed for $N=40$ and $T=48$~K, where the data are converged with respect to both system size and temperature. The $S(\mathbf{Q})$ values in (\textbf{G}) were extrapolated to the thermodynamic limit (see Appendix~\ref{app:QMC}). Panel (\textbf{H}) was calculated for $N=64$, where the crystal is commensurate with square boundary conditions and exhibits several higher-order correlation peaks.}
  \label{fig:stg_crystal}
\end{figure*}

To stabilize a spatially ordered state, we suppress the exciton zero-point motion by considering a larger mass, $m_X=5 m_0$, and higher densities. From an experimental perspective, such mass values are beyond those of currently studied trilayer exciton systems. However, enhanced masses can be realized via band structure engineering or moiré structures formed by a finite interlayer twist angle~\cite{Guo_2022,Brem_2020,Hafezi_2024,Malic_2020,ZhengTrilayerMoire2023,ChenBiTriMoire2022}.  
As we detail below, sufficiently strong dipolar interactions are only achieved at densities beyond the Mott transition in current systems, for the above choice of exciton mass. In this regime, excitons overlap and can no longer be treated as point bosons. Nevertheless, this extended parameter regime provides a more complete description of the phase diagram, which may prove relevant to future experimental systems. With this in mind, we consider an exciton density of $n=1.4\cdot 10^{14}$~cm$^{-2}$ and fix the dipole flipping rate $\Delta=26$~meV, as before. In addition, we carry out our calculations at temperatures corresponding to the ground-state limit of the system, as discussed in Appendix~\ref{app:QMC}. 

In Fig.~\ref{fig:stg_crystal}A, we depict a typical snapshot of the spatial world-line configuration. Indeed, we observe a staggered square lattice pattern, alternating between the two dipole orientations. The staggered crystal spontaneously breaks both translational and the Ising ($\mathbb{Z}_2$) dipole orientation symmetries of Eq.~\ref{eqn:H}. 
To provide a more quantitative measure, we compute the total structure factor,
\begin{equation} \label{eqn:Sk_def}
    S(\mathbf{k}) = \frac{1}{N} \expval{ \left|\sum_{i=1}^{N} e^{i\mathbf{k}\cdot\mathbf{r}_{i}}\right|^2} \, ,
 \end{equation}
and its dipole moment resolved counterpart,
 \begin{equation} \label{eqn:Sk_sigma}
    S_{\sigma}(\mathbf{k}) = \frac{1}{N}\expval{ \left|\sum_{i=1}^{N_\sigma} e^{i\mathbf{k}\cdot\mathbf{r}_{i}}\right|^2} \, ,
 \end{equation}
where the sum runs over the $N_\sigma$ excitons with dipole moment pointing in the direction $\sigma_z=+1$.

Figures~\ref{fig:stg_crystal}B and \ref{fig:stg_crystal}C present two-dimensional momentum-space maps of $S(\mathbf{k})$ and $S_\sigma(\mathbf{k})$, respectively. We find that $S(\mathbf{k})$ develops pronounced Bragg peaks at $\mathbf{k}=\mathbf{Q}=\frac{2\pi }{a}\hat{x}/\hat{y}$, where $a=1/\sqrt{n}$, consistent with a square lattice order. Simultaneously, the momentum map of $S_\sigma(\mathbf{k})$ peaks at a distinct Bragg vector $\mathbf{k}=\mathbf{Q}'=\frac{\pi}{a}\left(\hat{y}\pm\hat{x}\right)$, corresponding to an enlarged lattice constant that is also rotated by $\pi/4$ relative to $\mathbf{Q}$. Taken together, these results suggest the formation of a staggered dipolar square lattice that remains stable under zero-point motion, which, as detailed in Appendix~\ref{app:QMC}, we find to be significant~\cite{Gazit_2016}.

In Appendix~\ref{app:QMC}, we study the stability of the staggered crystal to density variations. We find that the crystal phase persists within the density range  $5.5\cdot10^{13}$~cm$^{-2}$$\leq n \leq5.5\cdot10^{14}$~cm$^{-2}$ beyond which it melts into superfluid phases at low and high densities. 

Focusing on the density region of a stable crystal, we now investigate the evolution of the exciton energy shifts, which may provide experimental fingerprints for the crystal phase. We first determine the scaling of $\mu_X$ with exciton density. The results, presented in Fig.~\ref{fig:stg_crystal}D, are consistent with a power-law behavior, $\mu_X \propto n^{\gamma}$, with $\gamma=1.40\pm 0.03$, as extracted from a linear fit to the data. This scaling law can be traced to the average electrostatic energy per exciton in the staggered lattice $E_{stg}$. At low densities, dipolar interactions scale as $r^{-3}$, such that $E_{stg}\propto n^{3/2}$~\cite{Slobodkin_2020}. The good agreement between our result and the above-estimated scaling of the electrostatic energy highlights the central role of dipolar interactions in the crystal and provides a direct experimental probe for this phase.

Next, we consider the stability of the crystal under an externally applied out-of-plane electric field. We initially examine the exciton chemical potential and find a nontrivial behavior as a function of $E_z$, as can be seen in Fig.~\ref{fig:stg_crystal}F. When increasing $E_z$ from the unbiased case, $\mu_X$ exhibits a redshift at weak field values. In the weak external field limit, as opposed to the liquid phases studied above, the dipole orientations in the staggered crystal are pinned by the lattice and are not susceptible to the addition of a single exciton. Therefore, the energy cost associated with dipolar interactions between the staggered lattice and an added exciton is insensitive to the field strength. Consequently, the observed redshift primarily originates from the standard dipolar coupling of the additional exciton to the external electric field. 

At stronger electric fields, however, this trend reverses, and $\mu_X$ displays an energy blueshift, indicating increasingly repulsive interactions. To better understand this behavior, we examine the evolution of the total structure factor and superfluid stiffness as a function of the electric field, as summarized in Fig.~\ref{fig:stg_crystal}G. We find that the Bragg peak $S(\mathbf{Q})$ vanishes and $\rho_s$ rises at the same electric field value where the redshift in $\mu_X$ turns into a blueshift. This correlated behavior directly points to the melting of the crystal into a superfluid phase. 

As detailed in Appendix~\ref{app:QMC}, this electric-field-induced melting coincides with the onset of dipole moment polarization, reflected in a finite average dipole moment per exciton. This mechanism is further illustrated by the evolution of spatial exciton correlations with the external electric field. In Fig.~\ref{fig:stg_crystal}H, we present the results for the parallel component of the spatial correlation function, $g_{\uparrow\uparrow}(r)$, corresponding to dipoles aligned with the field. The behavior of the remaining components of $g_{\sigma\sigma'}$ is discussed in Appendix~\ref{app:QMC}. At zero bias, the primary peak of $g_{\uparrow\uparrow}(r)$ corresponds to the nearest-neighbor distance between parallel dipoles in the staggered crystal, $\sqrt{2}a$. Increasing $E_z$ shifts the peak to shorter distances until it coincides with the mean interexciton separation. Concurrently, the weight of parallel dipolar configurations aligned with $E_z$ grows, as evident in the enhanced amplitude of $g_{\uparrow\uparrow}(r)$. Together, these results demonstrate the melting of the crystal, driven by the polarizing effect of the external electric field.

At still higher electric fields, enhanced dipolar repulsion of gradually polarized excitons results in a blueshift of $\mu_X$, similarly to the trend observed in the liquid phase above (see Fig.~\ref{fig:QP_n_evol}A). Finally, in the large $E_z$ limit, $\mu_X$ recovers the expected linear redshift at complete polarization, signaling the formation of a dipolar superfluid phase. In Fig.~\ref{fig:stg_crystal}E, we depict this state by displaying a typical QMC world-line configuration snapshot.

We note that the above nontrivial trend in $\mu_X$ provides a signature for the staggered lattice phase. In particular, the linear redshift at weak fields is unique to the crystal, as it stems from its stability against polarization, which is absent in the liquid phase (e.g. Fig.~\ref{fig:QP_n_evol}A).

Finally, in Appendix~\ref{app:QMC}, we analyze the onset of a staggered crystal phase as the exciton mass increases from the typical experimental value $m_X=m_0$. We find that a crystal appears at exciton masses $m_X \gtrsim 4m_0$.

\section*{Summary and Discussion}

We have established the critical role of attractive antiparallel dipolar interactions in shaping quantum collective phenomena in trilayer exciton structures. Specifically, we have demonstrated that antiparallel correlations are enhanced by suppressing charge-tunneling rates, increasing exciton density, or both. The resulting phase diagram exhibits a rich structure away from the quadrupolar limit (Fig.~\ref{fig:trilayer_overview}, C and D), including a self-bound droplet state, a partially fragmented condensate, and a staggered dipolar crystal. These phases show distinct dependencies of the exciton energy shifts on the out-of-plane electric field and density, providing insights into their driving mechanisms.

In that regard, we discuss our findings in light of recent experimental observations in TMD trilayers. Much of the experimental effort has focused on understanding the dependence of exciton energy shifts on the out-of-plane electric field at high exciton densities. Our results indicate a continuous crossover from a quadratic (quadrupolar) redshift in the dilute limit to a blueshift trend at high densities. This emergent blueshift arises from the competition between antiparallel attractive dipolar interactions and the progressive polarization of the dipoles as the electric field increases. A similar evolution of the energy shifts was observed in several recent experiments in trilayer excitons ~\cite{Yu_2023,Li_2023,Bai_2023,Lian_2023,Xie_2023,Xie_2024,Meng_2025}. Our results therefore highlight the importance of strong dipolar interactions and the formation of antiparallel configurations in interpreting the experimental observations.  

While stabilizing and detecting a staggered dipolar crystal phase in exciton trilayers is a key experimental challenge, our numerical results suggest that achieving this phase requires a relatively large exciton mass and high densities near the Mott transition threshold. Thus, this phase remains beyond the scope of current experiments. Nevertheless, should future experimental advances enable access to this high-mass regime, we provide a distinct experimental fingerprint of the crystal phase by examining its evolution under an external out-of-plane electric field. Concretely, the exciton energy shifts exhibit a linear redshift near zero electric field, which is a precursor to the electric-field-induced melting of the staggered lattice into a superfluid state.

Our prediction of a stable exciton droplet at experimentally accessible exciton densities suggests that near‑future experiments in TMD trilayers with increased interlayer spacing may directly probe this state. To provide quantitative guidance, we estimate the typical size of a single exciton droplet. Considering exciton densities in the range $n=10^{10}{-}10^{12}$~cm$^{-2}$ and a laser spot diameter of $w=1~\mu$m, the total number of generated excitons is then $N=n\pi (w/2)^2$. The local density within the droplet, $n_{\text{drop}}=3.2\cdot 10^{13}$~cm$^{-2}$, is determined by the numerically obtained mean interexciton distance, $a_{\text{drop}}=1.75$~nm. With these values, we estimate the droplet diameter to be $w_{\text{drop}}=17.5{-}175$~nm for the considered density range. $w_{\text{drop}}$ lies below the diffraction limit of typical optical setups, implying that the droplet is closely packed. Nevertheless, the droplet phase can be experimentally detected through the exciton emission profile. Under laser excitation of diameter $w$, the initially spread exciton cloud is expected to shrink as the excitons form a droplet, leading to a spatial emission spot in the sub-wavelength limit.

Although our phase diagram is mapped at zero temperature, the thermal stability of the staggered dipolar crystal can be estimated via the anisotropic generalization of the Kosterlitz-Thouless-Halperin-Nelson-Young (KTHNY) melting condition \cite{Podolsky_melting_2026}, $16\pi = \frac{Ya^2}{k_B T_m\sqrt{1+\zeta/4}}$, 
where $Y$ is the two-dimensional Young's modulus and $\zeta$ is the anisotropy parameter. For the crystal density, $n=1.4\cdot 10^{14}$ cm$^{-2}$, we obtain a melting temperature of $T_m\approx 178$ K, as detailed in Appendix~\ref{app:QMC}. We note that this estimate relies on a harmonic lattice model. The significant zero-point motion in the crystal (see Appendix~\ref{app:QMC}) is expected to introduce anharmonic corrections to the phononic dynamics beyond our approximation and consequently shift the melting temperature.

Obtaining a similar estimate for the thermal stability of the droplet requires a detailed computational study involving the characterization of its finite-temperature behavior. Given the high computational demands of this task, we leave it for future work.

We conclude our discussion by outlining several future research directions pertinent to our work. In a recent study~\cite{Meng_2025}, the introduction of a moiré-induced triangular external potential led to the formation of a confined exciton pair with a locally staggered configuration in each Moiré cell. Resolving the resulting phase diagram and the experimental signatures of trilayer excitons in an external periodic potential remains an open question that can be addressed using methods similar to those applied in our work. 

An additional intriguing direction concerns multilayered structures that extend beyond the trilayer limit and comprise multiple electron and hole layers. Here, the additional layer indices can allow for complex dipolar configurations that cannot be captured solely by dipole orientation fluctuations. Specifically, in this setting, a relative vertical shift between dipoles can form. This, in turn, may drive novel broken symmetry patterns that generalize the staggered crystal state.

Lastly, recent experiments have highlighted the role of electron-exciton interactions in doped trilayer systems~\cite{Meng_2025,Xie_2024,Lian_2023,Knap_2024,Zerba_2024,Zerba_2025}. In particular, such strong correlations can induce a quadrupolar-to-dipolar transition distinct from the density-induced evolution presented in this work. Providing a microscopic model for this effect and establishing the resulting phase diagram are outstanding questions, which we leave to future research.

% \vfill
\begin{acknowledgments}
S.G. acknowledges support from the Israel Science Foundation (ISF) Grant no. 586/22. R.R. acknowledges support from the Israel Science Foundation (ISF) Grant no. 1087/22. D.P. acknowledges support from the Israel Science Foundation (ISF) Grant no. 2005/23. M.Z. acknowledges the support of the Council for Higher Education Scholarship Program for Outstanding Doctoral Students in Quantum Science and Technology.
\end{acknowledgments}

\section*{Data availability}
The data that support the findings of this article are openly available on Zenodo \cite{trilayer_zenodo}.

\onecolumngrid
\appendix
\section{QMC simulation \label{app:sim}}

In this appendix, we provide additional details on the QMC algorithm and the simulations used to study the trilayer system. We employ the path-integral QMC method \cite{Ceperley_1995} and implement the efficient continuous-space worm algorithm \cite{Boninsegni_2006}. Unique to our model are the additional dipole moment degrees of freedom, which require a nontrivial extension of standard methodologies. The dipole orientation is represented by an Ising variable, $\sigma^z$. To describe the dynamical flipping of dipole orientations, we devise an updating scheme that combines world-line deformations with dipole-flipping updates, as detailed below.

\subsection*{Path integral formulation}

We write the Hamiltonian in Eq.~1 as the sum:
\begin{equation} 
    \hat{H} = \hat{T} + \hat{U} + \hat{H}_{\sigma} \,,
\end{equation}
Here, $\hat{T}$ is the kinetic energy operator, $\hat{U}$ is the interexciton interaction term in Eq.~2, and $\hat{H}_{\sigma}$ governs the dynamics of the dipole moment orientation. Explicitly, 
\begin{equation}
    \hat{H}_{\sigma} = -\Delta \sum_{i=1}^N \hat{\sigma}^x_i - edE_z \sum_{i=1}^N \hat{\sigma}^z_i \,,
\end{equation}
where $N$ is the number of particles and $\hat{\sigma}^{x,z}_i$ are Pauli operators acting in the Ising subspace of the $i$th particle. In constructing a path integral representation of the partition function, we choose to work in the eigenbasis $\ket{\bar{\sigma}^z,\bar{R}}$, where $\ket{\bar{\sigma}^z} = \ket{\sigma^z_1,\sigma^z_2,...,\sigma^z_N}$ and $\ket{\bar{R}} = \ket{\mathbf{r}_1,\mathbf{r}_2,...,\mathbf{r}_N}$ are the collective Ising and position coordinates, respectively. This yields the partition function
\begin{equation} \label{eqn:Z}
    \mathcal{Z} = \int d \bar{R}\sum_{\{\ket{\bar{\sigma}^z}\}}\mel{\bar{\sigma}^z,\bar{R}}{e^{-\beta( \hat{T} + \hat{U} + \hat{H}_{\sigma})}}{\bar{\sigma}^z,\bar{R}} \, .
\end{equation}

Using the Trotter decomposition with $\epsilon = \beta/M$, we obtain 
\begin{equation}
    e^{-\epsilon( \hat{T} + \hat{U} + \hat{H}_{\sigma})} = e^{-\epsilon\hat{T}} e^{-\epsilon \hat{U}} e^{-\epsilon \hat{H}_{\sigma}} +\mathcal{O}(\epsilon ^2) \, ,
\end{equation}
which results in the path-integral representation

\begin{equation} \label{eqn:Z_expand}
    \mathcal{Z} = \int \left( \prod_{\tau=0}^{M-1} d\bar{R}_{\tau}\right) \sum_{\{\ket{\bar{\sigma}^z_{\tau}}\}} \prod_{\tau=0}^{M-1} \mel{\bar{\sigma}^z_{\tau+1}, \bar{R}_{\tau+1}}{e^{-\epsilon\hat{T}} e^{-\epsilon \hat{U}}e^{-\epsilon\hat{H}_{\sigma}}}{\bar{\sigma}^z_{\tau}, \bar{R}_{\tau}} \,.
\end{equation}
Here, $\tau$ denotes the imaginary time index, with periodic boundary conditions applied, $\ket{\bar{\sigma}^z_{M}, \bar{R}_{M}} = \ket{\bar{\sigma}^z_{0}, \bar{R}_{0}}$.

For an interaction potential operator that depends on both spatial and Ising variables, we have $e^{-\epsilon \hat{U}}\ket{\bar{\sigma}^z_{\tau}, \bar{R}_{\tau}} = e^{-\epsilon U(\bar{R}_{\tau},\bar{\sigma}^z_{\tau})}$. Thus, the matrix element in Eq.~\ref{eqn:Z_expand} factorizes into two components~\cite{Ceperley_1995,Landau_2021}: 
\begin{align}
    \mel{\bar{R}_{\tau+1}}{e^{-\epsilon \hat{T}} }{\bar{R}_{\tau}} &= \left( \frac{m}{2\pi\hbar^2\epsilon}\right)^{\frac{ND}{2}} e^{-\frac{m}{2\hbar^2\epsilon}\left(\bar{R}_{\tau+1} - \bar{R}_{\tau}\right)^2}, \\
    \mel{\bar{\sigma}^z_{\tau+1}}{e^{-\epsilon \hat{H}_{\sigma}}}{\bar{\sigma}^z_{\tau}} &= Q^N e^{K_\tau \sum_i \sigma^z_{i,\tau}\sigma^z_{i,\tau+1} \,-\, \epsilon edE_z \sum_i \sigma^z_{i,\tau}} \, .
\end{align}
The first term is a normalized Gaussian distribution, with $m$ the particle mass and $D$ the spatial dimension. The factors in the second term are given by
\begin{equation}
    Q = \sqrt{\sinh{(\epsilon \Delta)}\cosh{(\epsilon \Delta)}} \,, \, K_\tau = -\frac{1}{2}\ln{\left(\tanh{(\epsilon \Delta)}\right)} \, .
\end{equation}
Substituting the above expressions into Eq.~\ref{eqn:Z_expand} yields the final form, 
\begin{equation} \label{eqn:full_Z}
    \mathcal{Z} = \left( \frac{m}{2\pi\hbar^2\epsilon}\right)^{\frac{MND}{2}} Q^{MN} \int \left( \prod_{\tau=0}^{M-1} d\bar{R}_{\tau}\right) \sum_{\{\bar{\sigma}^z_{\tau}\}} e^{\sum_\tau \left[-\frac{m}{2\hbar^2\epsilon}\left(\bar{R}_{\tau+1} \,-\, \bar{R}_{\tau}\right)^2 -\epsilon U(\bar{R}_{\tau},\bar{\sigma}^z_{\tau}) \,+\, K_\tau\sum_i \sigma^z_{i,\tau}\sigma^z_{i,\tau+1} \,-\, \epsilon edE_z\sum_i \sigma^z_{i,\tau}\right]} \,.
\end{equation}

\subsection*{Sampling}

The above partition function is sampled using the path-integral Monte Carlo method \cite{Ceperley_1995}, with a continuous-space worm algorithm~\cite{Boninsegni_2006} that is generalized to include updates of the Ising variables.  

The configuration space of $\mathcal{Z}$, referred to as \textit{diagonal}, consists of particle world-lines. Each world-line comprises beads that represent the particle's location and Ising variable at a given imaginary-time slice. Each bead $\alpha$ is connected to the beads residing at its previous and next imaginary time slices, accessed by the functions $\text{prev}(\alpha)$ and $\text{next}(\alpha)$, respectively. Following the standard worm algorithm framework, the configuration space is extended to include off-diagonal configurations that contain a single open world-line, with endpoints denoted by Masha ($\mathcal{M}$) and Ira ($\mathcal{I}$). 

Since the observables of interest are diagonal, measurements are performed only in the diagonal sector. Off-diagonal configurations enable efficient sampling of bosonic permutations. The simulation efficiency can be optimized by tuning either the relative weight between the diagonal and off-diagonal sectors $C$, or the typical length of the worm update $\bar{M}$. See Ref.~\cite{Boninsegni_2006} for a detailed explanation.

To further facilitate the sampling of diagonal observables, we implement the worm algorithm in the canonical ensemble\footnote{To sample the chemical potential, we extend our canonical scheme to include grand canonical updates within a restricted configuration space, following the method proposed in Ref.~\cite{Del_Maestro_2014}.}. To maintain a fixed particle number, the worm updates are adapted so that the endpoints $\mathcal{M}$ and $\mathcal{I}$ always reside at the same imaginary-time slice. Within this canonical scheme, the diagonal sector is sampled using five complementary updates: Open, Close, Add, Delete, and Swap.

The spatial coordinates are sampled from the free-particle imaginary-time propagator,
\begin{equation}
    \rho(\mathbf{r},\mathbf{r}',\epsilon)=\left(\frac{m}{2\pi\hbar^2\epsilon}\right)^{\frac{D}{2}}\exp\left(-\frac{m}{2\hbar^2\epsilon}(\mathbf{r}-\mathbf{r}')^{2}\right) \,,
\end{equation}
associated with the kinetic energy cost of connecting two beads separated by an imaginary time step $\epsilon$. Given a starting point $\mathbf{r}$ of a segment of length $\tilde{M}$ time slices, the position of the endpoint $\mathbf{r}'$ is sampled from $\rho(\mathbf{r},\mathbf{r}',\epsilon\tilde{M})$. The intermediate beads are drawn using the multilevel sampling method described in \cite{Ceperley_1995}. 

The Ising variables are sampled from the local Ising interaction weight, which assigns a conditional probability to drawing an Ising value $\sigma'_\tau$ given $\sigma_{\tau-1}$ or vice versa, effectively determining the locations of temporal domain walls. The corresponding probability function reads
\begin{equation}\label{eqn:P_sigma}
    P_{\sigma}=\frac{e^{K_{\tau}\sigma_{\tau-1}\sigma'_{\tau}}}{2\cosh{K_{\tau}}} \, ,
\end{equation}
where we have dropped the $z$ superscript of $\sigma^z_\tau$ for brevity. For a segment with a fixed starting point, the Ising variables along the segment are generated sequentially using $P_\sigma$, propagating toward the endpoint. We note that this step is performed before sampling the spatial component of the world-line configurations. In certain cases where the endpoint Ising variable is fixed, the above-outlined sampling scheme leaves an additional Boltzmann weight associated with the imaginary-time Ising coupling. As detailed below, this is accounted for by reweighting the Metropolis-Hastings acceptance probability.

\textbf{Open.} Starting from a diagonal configuration, the Open move creates a worm as follows. A bead $\alpha$ is selected at random among the $MN$ beads. An integer $\tilde{M}\in [1,\bar{M}]$ is drawn from a uniform distribution, and a new bead $\tilde{\alpha}$ is placed $\tilde{M}$ time slices behind $\alpha$ in imaginary time. A new world-line path is generated between $\tilde{\alpha}$ and $\alpha$. Ising variables are sampled sequentially using $P_\sigma$, starting from the known Ising value of $\alpha$ and propagating backward in imaginary time. The spatial coordinates are drawn from $\rho$ using multilevel sampling. Then, the existing path between $\alpha$ and $\text{prev}^{\tilde{M}}(\alpha)$ is deleted, and the worm endpoints are set to $\mathcal{I}=\text{prev}^{\tilde{M}}(\alpha)$, $\mathcal{M}=\tilde{\alpha}$. The acceptance probability for this move is given by
\begin{equation} \label{eqn:A_open}
    A_{\text{open}} = min\left(1,\frac{CMN}{\rho(r_{\tilde{\alpha}},r_{\alpha},\epsilon \tilde{M})} \cdot \frac{2\cosh(K_\tau)}{e^{K_\tau \sigma_1\sigma_2}}e^{-\epsilon\left(\Delta U+\Delta U_{z}\right)} \right) \, ,
\end{equation}
where $\Delta U = U(\bar{R}',\bar{\sigma}') - U(\bar{R},\bar{\sigma})$ is the change in the dipolar energy and $\Delta U_z = edE_z\sum_{\tau \in \mathcal{P}}(\sigma'_\tau - \sigma_\tau)$ is the difference in the electric-field-induced energy along the set $\mathcal{P}$ of modified time slices. The Ising values $\sigma_1=\sigma_{\text{prev}^{\tilde{M}}(\alpha)}$ and $\sigma_2=\sigma_{\text{prev}^{\tilde{M}-1}(\alpha)}$ are given by the initial configuration. The remaining prefactors fulfill detailed balance with respect to the complementary Close move, as discussed below.

\textbf{Close.} This move removes a worm and transitions back to the diagonal sector.  An integer $\tilde{M}\in [1,\bar{M}]$ is chosen uniformly, and a bead $\alpha=\text{next}^{\tilde{{M}}}(\mathcal{M})$ is defined. A new world-line segment from $\mathcal{I}$ to $\alpha$ is proposed by first generating the Ising values, starting from $\alpha$ and propagating backward in imaginary time down to time slice $\tau_{\text{prev}^{\tilde{M}-1}(\alpha)}$. The spatial coordinates are then sampled from $\rho$ in a multilevel manner. The existing segment between $\mathcal{M}$ and $\alpha$ is deleted, and the worm endpoints are removed. The corresponding acceptance probability for Close reads
\begin{equation} \label{eqn:A_close}
    A_{\text{close}} = min\left(1,\frac{\rho(r_{\mathcal{I}},r_{\alpha},\epsilon \tilde{M})}{CMN} \cdot \frac{e^{K_\tau \sigma_1\sigma'_2}}{2\cosh(K_\tau)}e^{-\epsilon\left(\Delta U+\Delta U_{z}\right)} \right) \, .
\end{equation}
Here, $\sigma_1=\sigma_{\mathcal{I}}$ is known from the initial configuration, while $\sigma'_2=\sigma'_{\text{prev}^{\tilde{M}-1}(\alpha)}$ is the last generated Ising value along the proposed world-line segment. 

The prefactors in Eqs.~\ref{eqn:A_open} and \ref{eqn:A_close} ensure detailed balance between the Open and Close moves. The factor $C$ originates from the transition between the off-diagonal and diagonal sectors. The term $MN$ accounts for the selection of the bead $\alpha$ during the Open move, while in the Close move this bead is predetermined. $\rho$ and the Ising weight arise from the proposal probabilities of the spatial and Ising values of the bead $\tilde{\alpha}$ in Open, whereas in Close this bead corresponds to $\mathcal{I}$.

\textbf{Advance forward (AF).} The AF move advances the worm endpoints in imaginary time. A random integer $\tilde{M}\in [1,\bar{M}]$ is chosen, and a bead $\alpha=\text{next}^{\tilde{M}}(\mathcal{I})$ is defined. A new path from $\mathcal{I}$ to $\alpha$ is generated using the same sequential Ising sampling, starting from $\mathcal{I}$, and using multilevel spatial coordinate sampling as in the Open move. Then, the path between $\mathcal{M}$ and $\tilde{\alpha}=\text{next}^{\tilde{M}}(\mathcal{M})$ is deleted, and the worm endpoints are updated as $\mathcal{I}=\alpha, \mathcal{M}=\tilde{\alpha}$. The move is accepted with probability
\begin{equation} \label{eqn:A_add}
    A_{\text{AF}} = min\left(1,e^{-\epsilon\left(\Delta U+\Delta U_{z}\right)} \right) \, ,
\end{equation}
which depends only on the changes in the dipolar interaction and in electric-field-induced energies, since proposal probabilities cancel between the AF and its reverse move, Advance Backward, which is described below. 

\textbf{Advance backward (AB).} AB is the time-reversed counterpart of AF, and propagates the worm endpoints backward in imaginary time. A random integer $\tilde{M}\in [1,\bar{M}]$ is drawn, and a bead $\alpha=\text{prev}^{\tilde{M}}(\mathcal{M})$ is defined. A new segment is generated from $\mathcal{M}$ to $\alpha$, where Ising values are sequentially drawn from $P_\sigma$, propagating from $\mathcal{M}$ in the negative imaginary time direction, and using multilevel sampling of the spatial coordinates. Next, the segment connecting $\mathcal{I}$ and $\tilde{\alpha}=\text{prev}^{\tilde{M}}(\mathcal{I})$ is removed, and the worm endpoints are set to $\mathcal{I}=\tilde{\alpha}$ and $\mathcal{M}=\alpha$. The corresponding acceptance probability is
\begin{equation} \label{eqn:A_del}
    A_{\text{AB}} = min\left(1,e^{-\epsilon\left(\Delta U+\Delta U_{z}\right)} \right) \, .
\end{equation}

\textbf{Swap.} This move efficiently samples bosonic permutations by alternating between different topological sectors of off-diagonal configurations. A bead $\alpha$ is selected at time slice $\tau_{\text{next}^{\bar{M}}(\mathcal{I})}$, defining a second bead,  $\zeta=\text{prev}^{\bar{M}}(\alpha)$, along its imaginary-time path. A new path is generated from $\mathcal{I}$ to $\alpha$. Ising variables are sequentially sampled in the negative imaginary-time direction, starting from $\alpha$, and spatial coordinates are drawn using multilevel sampling. The original path connecting $\zeta$ and $\alpha$ is then removed, and the worm endpoint is updated as $\mathcal{I}=\zeta$. The Swap acceptance probability is given by
\begin{equation} \label{eqn:A_swap}
    A_{\text{swap}} = min\left(1,\frac{\Sigma_{\mathcal{I}}}{\Sigma_{\zeta}} \cdot e^{K_\tau\left( \sigma_\mathcal{I}\sigma'_{\text{next}(\zeta)}- \sigma_{\zeta}\sigma_{\text{next}(\zeta)}\right)} e^{-\epsilon\left(\Delta U+\Delta U_{z}\right)} \right) \, ,
\end{equation}
where the $\Sigma_{\mathcal{I}}/\Sigma_{\zeta}$ factor accounts for the selection probability of $\alpha$ in the two complementary Swap moves; see Ref.~\cite{Boninsegni_2006} for details.

\subsection*{Observables}

Our simulations provide access to several observables that characterize the different phases of the system. The total and dipole-orientation-resolved structure factors, as well as the spatial correlation function, are defined in the main text and depend explicitly on the spatial coordinates and are thus directly obtained from the sampled world-line configurations. These quantities are used to identify spatial correlations and ordering, characterizing crystalline or liquid phases.

The path-integral framework also enables the calculation of the superfluid stiffness $\rho_s$ through winding number statistics \cite{Ceperley_1995}. Specifically, in $D=2$ spatial dimensions, the corresponding estimator is 
\begin{equation}
    \rho_{s} = \frac{L_x^2\expval{W_x^2} + L_y^2\expval{W_y^2}}{A\beta D} \,,
\end{equation}
where $L_{x(y)}$ and $W_{x(y)}$ are the box size and the winding number in the $x\,(y)$ direction, $A$ is the area, and $\beta$ is the inverse temperature. This definition allows for rectangular simulation boxes, which are required for commensurability in the case of crystalline order. $\rho_{s}$ is expected to be finite in the superfluid phase and vanishes otherwise. An additional observable for detecting superfluidity is the dipole-orientation-resolved superfluid fraction, $n_{\sigma\sigma'}(k=0)$, defined in the main text, which serves as an order parameter for a single-particle condensate, providing information about the dipolar state of the exciton condensate and its fragmentation.

Beyond spatial correlations, the simulation also probes the dipole orientation degree of freedom. The mean dipole moment per particle is computed as
\begin{equation} \label{eqn:Stot}
    \expval{\sigma^{z}} = \frac{1}{N} \expval{ \sum_{i=1}^{N} \sigma_i^z} \,,
\end{equation}
which quantifies the net polarization induced by the external out-of-plane electric field $E_z$. In the absence of an external bias, the mean dipole moment vanishes because of the inherent $\mathbb{Z}_2$ symmetry of the Hamiltonian.
By contrast, a fully polarized configuration yields $\expval{\sigma^{z}}=1$.

Dipole moment fluctuations are computed via the imaginary-time dipole-orientation correlation function, 
\begin{equation}\label{eqn:G_tau}
    G(\tau)=\left\langle \sigma_i^z(0)\sigma_i^z(\tau)\right\rangle \,.
\end{equation}
$G(\tau)$ allows the determination of the quantum nature of the quadrupolar phase, characterized by rapid fluctuations of the dipole orientation. In this case, the imaginary-time dynamics are expected to obey an exponential decay set by the effective dipolar-quadrupolar energy gap $\Delta_\text{DQ}$ \cite{Slobodkin_2020}. When varying the inverse temperature, the decay rate of these correlations scales as $G(\tau=\beta/2) \propto e^{-\Delta_\text{DQ}\beta}$.

\subsection*{Benchmarking}

To benchmark our simulation, we compare it to the exact solution of the trilayer model in the two-body limit. In the absence of an external out-of-plane electric field, the system is governed by the Schr\"{o}dinger equation,

\begin{equation} \label{eqn:H_2bd}
    \left[-\frac{\hbar^2}{2m_r }\nabla^2 +\hat{U}(r,\sigma^z_1,\sigma^z_2) -\Delta(\hat{\sigma}^x_1 + \hat{\sigma}^x_2) \right]\psi = E\psi \, ,
\end{equation}
where $r$ denotes the relative spatial coordinate, $m_r=0.5m_X$ is the reduced mass, and the interaction term is given by the dipolar potential defined in Eq.~2. Introducing the dipole length $d$ and the energy scale $\varepsilon = \frac{e^2}{\kappa d}$, we define the dimensionless variables $\tilde{r}=r/d$, $\tilde{\alpha} = \frac{\hbar^2 \kappa }{2m_r d e^2}$, $\tilde{\Delta}=\Delta/\varepsilon$, $\tilde{U}=U/\varepsilon$ and $\tilde{E} = E/\varepsilon$. In these units, the Schr\"{o}dinger equation becomes
\begin{equation}
   \left[-\tilde{\alpha} \tilde{\nabla}^2 +\hat{\tilde{U}}(\tilde{r},\sigma^z_1,\sigma^z_2) -\tilde{\Delta}(\hat{\sigma}^x_1 + \hat{\sigma}^x_2)\right]\psi = \Tilde{E}\psi \, ,
\end{equation}
where $\tilde{\nabla}$ denotes the gradient with respect to the dimensionless coordinate $\tilde{r}$. The exact solution is obtained using the improved exact-diagonalization method described in \cite{Laliena_2018}. 

\begin{figure}[h!]
    \centering
    \includegraphics[width=0.7\textwidth]{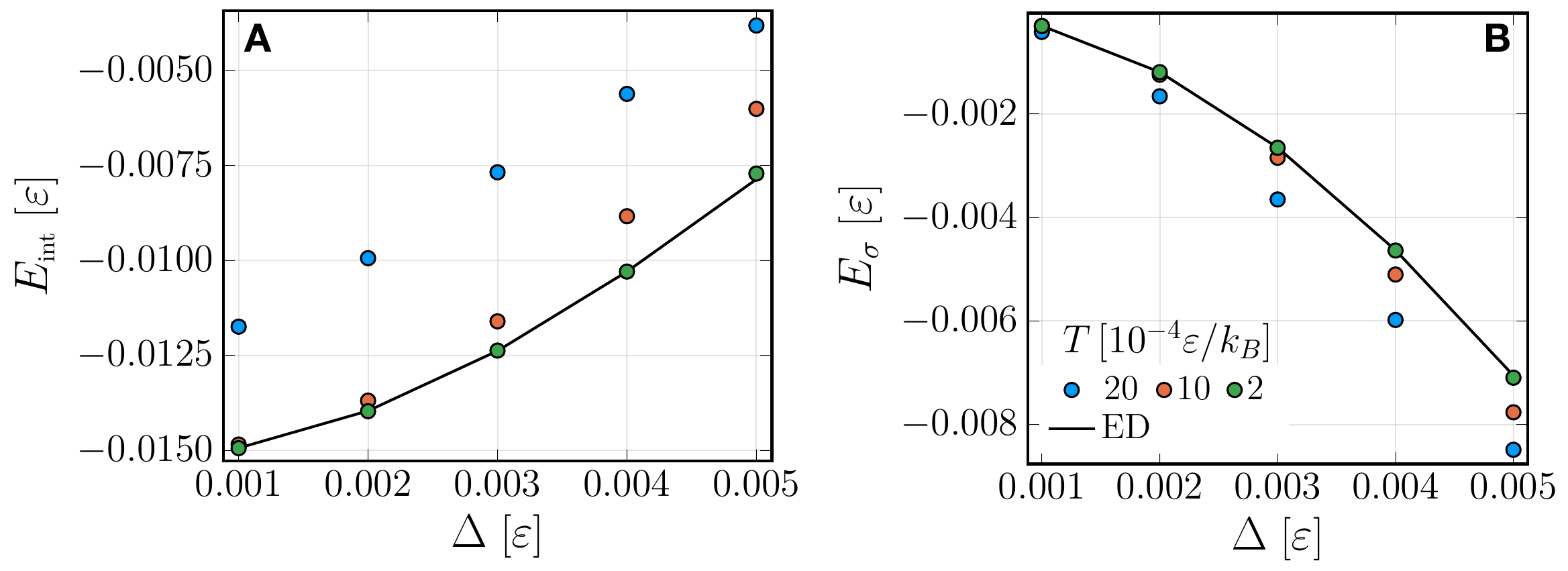}

    \caption{\textbf{Benchmarking the two-body trilayer model energies.} (\textbf{A}) Interaction energy and (\textbf{B}) dipole-flip energy as a function of the dipolar-quadrupolar energy gap, shown for three decreasing temperatures, as labeled in (\textbf{B}). The black line corresponds to the exact solution of the two-body problem.
      }
  \label{fig:Es_trilayer_ED}
\end{figure}

To validate our QMC implementation, we examine the convergence of the interaction and Ising energies, defined as $E_{\text{int}}=\expval{\hat{U}}$ and $E_{\sigma}=\expval{\hat{H}_\sigma}$, to their ground state values as the temperature decreases. As shown in Fig.~\ref{fig:Es_trilayer_ED}, the QMC results recover the ground-state values in the low-temperature limit across a range of $\Delta$ values. This agreement confirms the correct implementation of our QMC algorithm in both the Ising and interaction sectors.

\begin{figure}[ht!]
    \centering
    \includegraphics[width=0.8\textwidth]{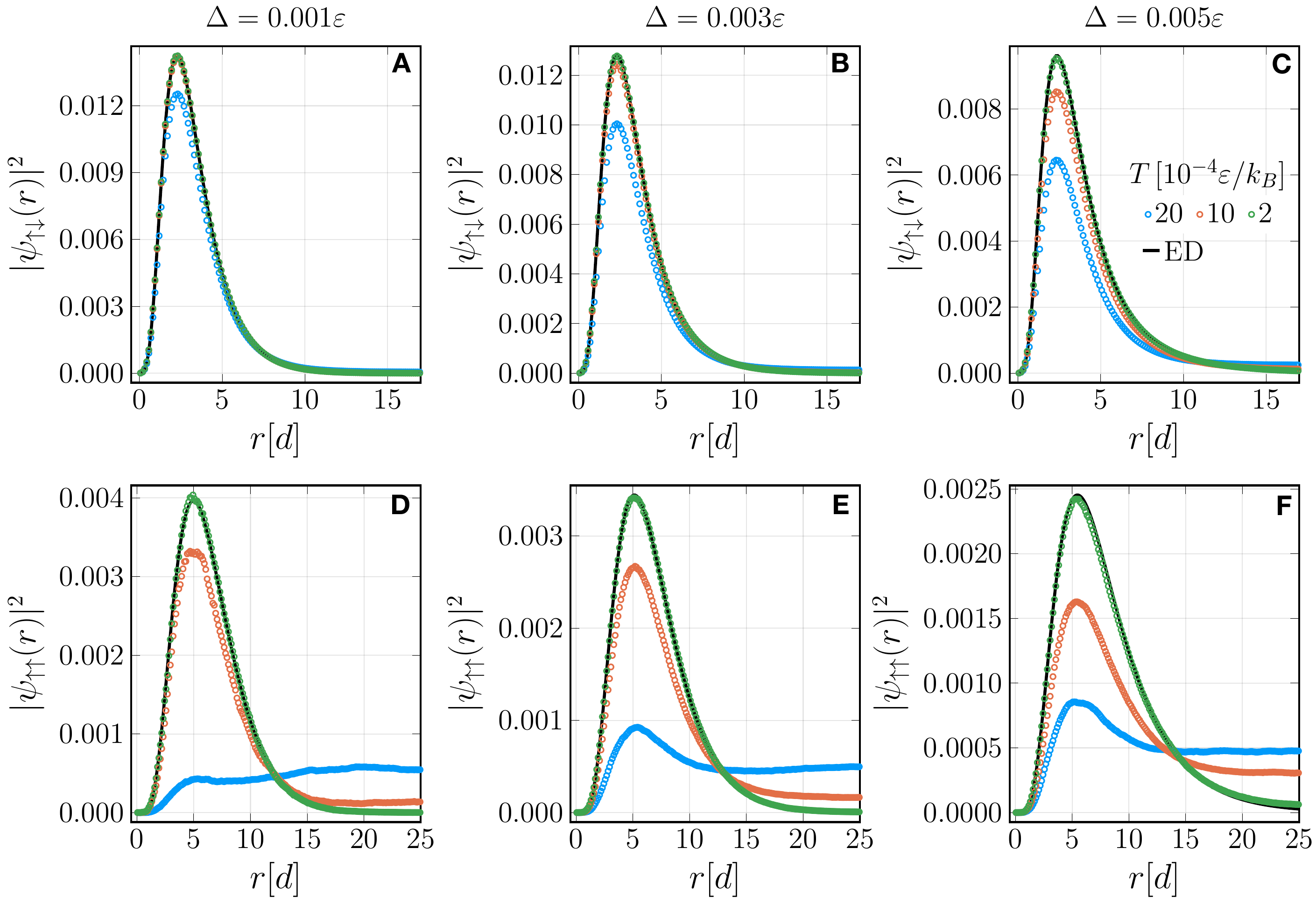}
    \caption{\textbf{Benchmarking the two-body trilayer model spatial probability distributions.} Panels (\textbf{A})-(\textbf{C}) show the spatial probability distribution function of antiparallel dipoles in the two-body limit of the trilayer model for dipolar-quadrupolar energy gaps $\Delta=\{0.001,0.003,0.005\}\varepsilon$, respectively. Colored data points are the QMC results for successively lower temperatures, as labeled in (\textbf{C}). The black line is the zero-temperature solution obtained from exact diagonalization. The corresponding results for the parallel component are presented in panels (\textbf{D})-(\textbf{F}). Each distribution function is normalized within its respective configurational subspace.
  }
  \label{fig:Psi_trilayer_ED}
\end{figure}

To further test the coupling between the spatial and Ising degrees of freedom, we evaluate the exciton spatial distribution functions in the parallel and antiparallel Ising subspaces, $\left| \psi_{\uparrow\uparrow}\right|^2$ and $\left| \psi_{\uparrow\downarrow}\right|^2$. The results are presented in Fig.~\ref{fig:Psi_trilayer_ED} for several values of $\Delta$. The QMC results reach the exact solution as the temperature decreases, further confirming the validity of our simulation.

\section{Additional QMC data \label{app:QMC}}

\subsection*{Quadrupolar superfluid}
\begin{figure}[ht!]
    \centering
    \includegraphics[width=0.7\textwidth]{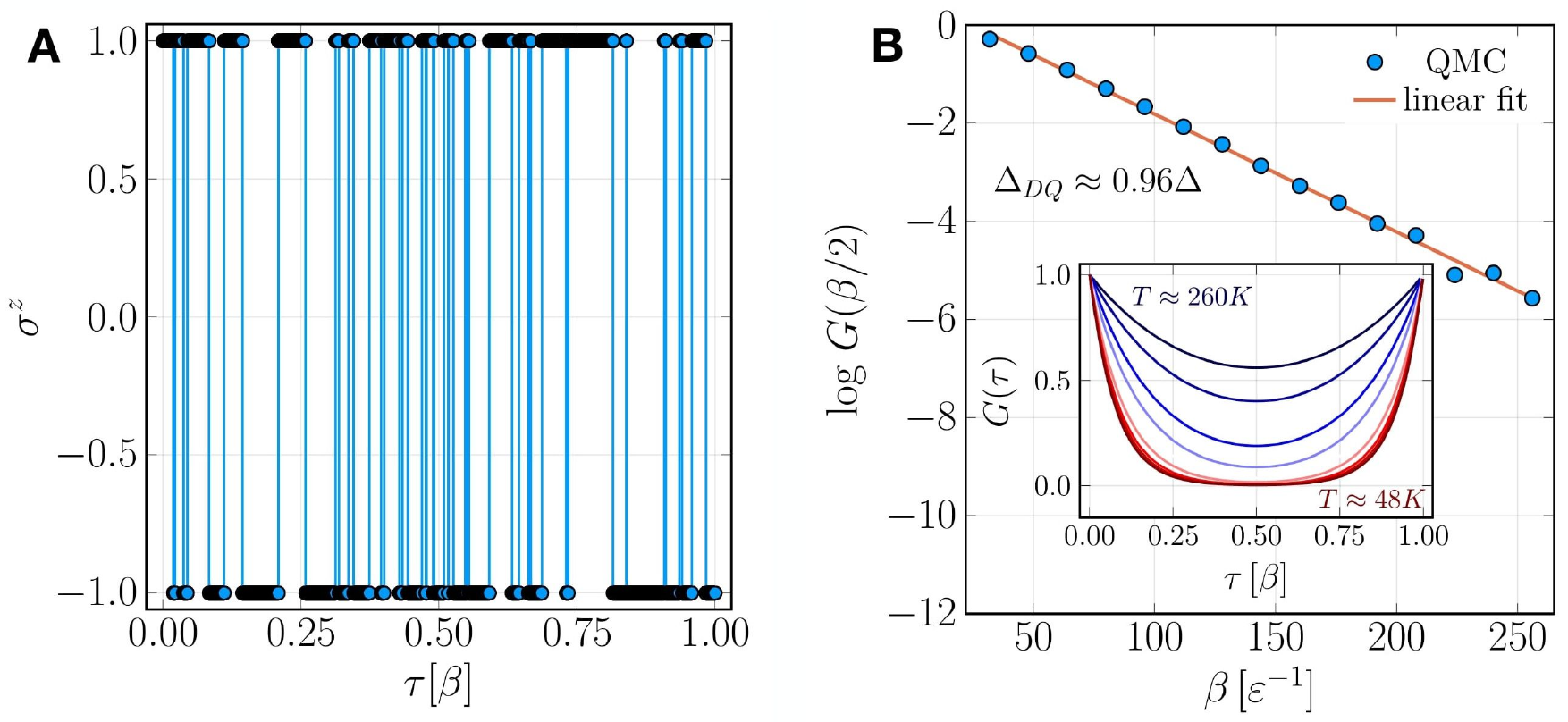}
    \caption{\textbf{The quantum nature of the quadrupolar state.} (\textbf{A}) The exciton dipole moment as a function of imaginary time, shown for $n=1.4\cdot 10^{12}$~cm$^{-2}$ and $T=6$~K. (\textbf{B}) The dipolar-quadrupolar energy gap is extracted from a semi-logarithmic plot of the dipole orientation correlation decay as a function of $\beta$. Blue circles are numerical data points corresponding to the value of the dipole orientation correlation function at maximal temporal distance $\tau=\beta/2$. The red line is a linear fit to the data, yielding a gap value close to the microscopic input $\Delta_\text{DQ}\approx0.96 \Delta_0$. Each data point was taken from the midpoints of $G(\tau)$ curves calculated for temperatures in the range $48$~K$\leq T\leq 260$~K, as shown in the inset. These data were obtained for $N=32$ and $T=193$~K.
  }
  \label{fig:quantum_QP}
\end{figure}

In this appendix, we provide an extended investigation of the quadrupolar phase at $m_X=m_0$ and $\Delta=26$~meV, beyond the numerical data presented in the main text. We first demonstrate the quantum nature of the quadrupolar state in the dilute exciton density limit of $n=1.4\cdot 10^{12}$~cm$^{-2}$. To this end, we examine the single dipole orientation correlation function as a function of imaginary time, $G(\tau)$. At low temperatures, the exciton dipole moment exhibits pronounced temporal fluctuations, as evident in Fig.~\ref{fig:quantum_QP}A, leading to an exponential decay of $G(\tau)$, as shown in the inset of Fig.~\ref{fig:quantum_QP}B. To quantify these fluctuations, we analyze $G(\tau)$ at the maximal imaginary time separation, $\tau = \beta/2$, as a function of inverse temperature. As shown in Fig.~\ref{fig:quantum_QP}B, we find a clear linear dependence of our data on a semilogarithmic scale. A linear fit yields an effective dipolar-quadrupolar energy gap of $\Delta_\text{DQ}=25.5\pm 0.4$~meV, which is in good agreement with the bare value. The above results establish the quantum nature of the quadrupolar state, borne from quantum fluctuations in the dipolar orientation between the outer layers.

\begin{figure}[h!]
    \centering
    \includegraphics[width=0.95\textwidth]{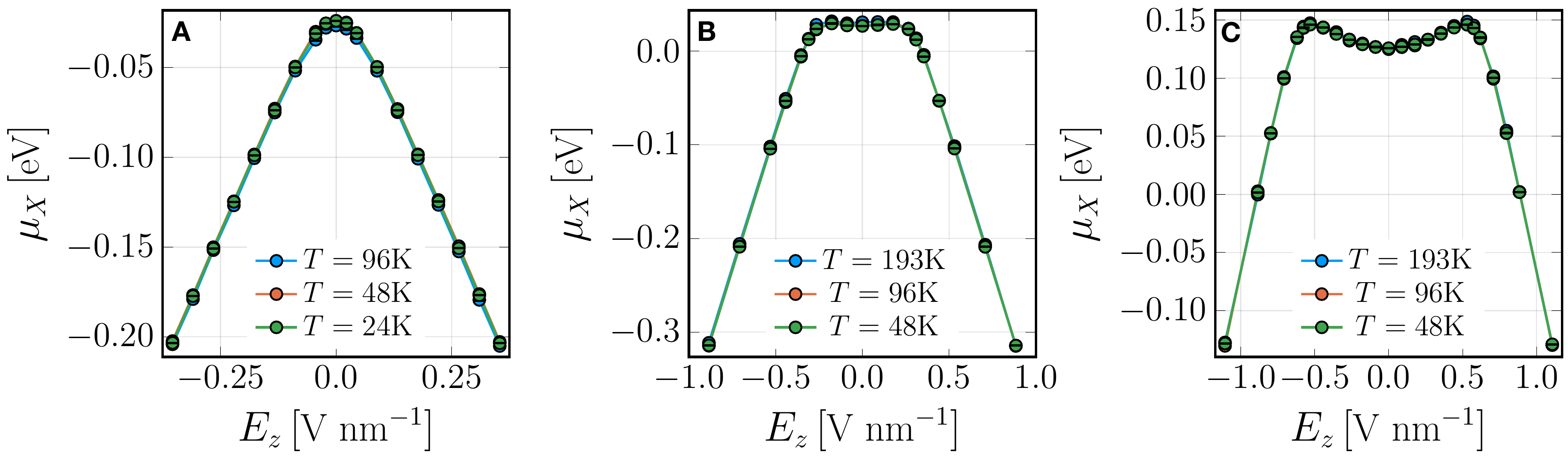}

    \caption{\textbf{Temperature convergence of the exciton chemical potential.} 
    The chemical potential is shown as a function of $E_z$ for a set of decreasing temperatures. Panels correspond to exciton densities of (\textbf{A}) $n=1.4\times10^{12}$~cm$^{-2}$, (\textbf{B}) $n=14\times10^{12}$~cm$^{-2}$, and (\textbf{C}) $n=28\times10^{12}$~cm$^{-2}$. The system size was set to $N=64$.
    }
  \label{fig:mu_Tconv}
\end{figure}

Next, we track the convergence of the exciton chemical potential under an out-of-plane electric field to the zero-temperature limit. In Fig.~\ref{fig:mu_Tconv}, we present $\mu_X(E_z)$ for decreasing temperatures at exciton densities $n=\{1.4$, $14$, $28\}\cdot 10^{12}$~cm$^{-2}$, as considered in Fig.~3A of the main text. For all densities, the results converge at $T=48$~K, verifying the low-temperature limit of our data.

\subsection*{Exciton droplet and partially fragmented superfluid}

Here, we present additional results for the exciton droplet and the partially fragmented superfluid observed at $\Delta=1$~meV. 
We begin by testing the ground-state convergence of our data within the droplet phase. To this end, we compute the antiparallel component of the exciton pair correlation function, $g_{\uparrow\downarrow}(r)$, across a range of decreasing temperatures. As shown in Fig.~10, curves at different temperatures overlap at short distances, demonstrating clear agreement on the location of the droplet's characteristic correlation peak. At longer distances, the curves show a converging trend as the temperature decreases. Based on these observations, we use $T=1.5$~K as the effective zero-temperature limit of our simulations.

\begin{figure}[ht!]
    \centering
    \includegraphics[width=0.36\textwidth]{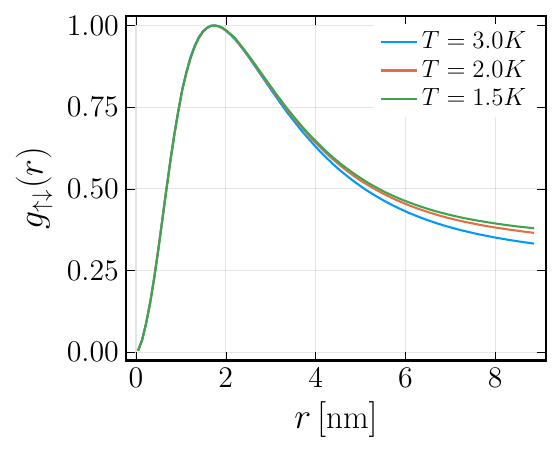}

    \caption{\textbf{Finite temperature convergence in the droplet state.} The exciton antiparallel correlation function in the droplet phase was computed for $T=$3,2 and 1.5 K. Here, the exciton density is $n=1.4\cdot 10^{11}$ cm$^{-2}$ and the particle number is set to $N=32$.
    }
    \label{fig:droplet_T_conv}
\end{figure}

\begin{figure}[ht!]
    \centering
    \includegraphics[width=0.65\textwidth]{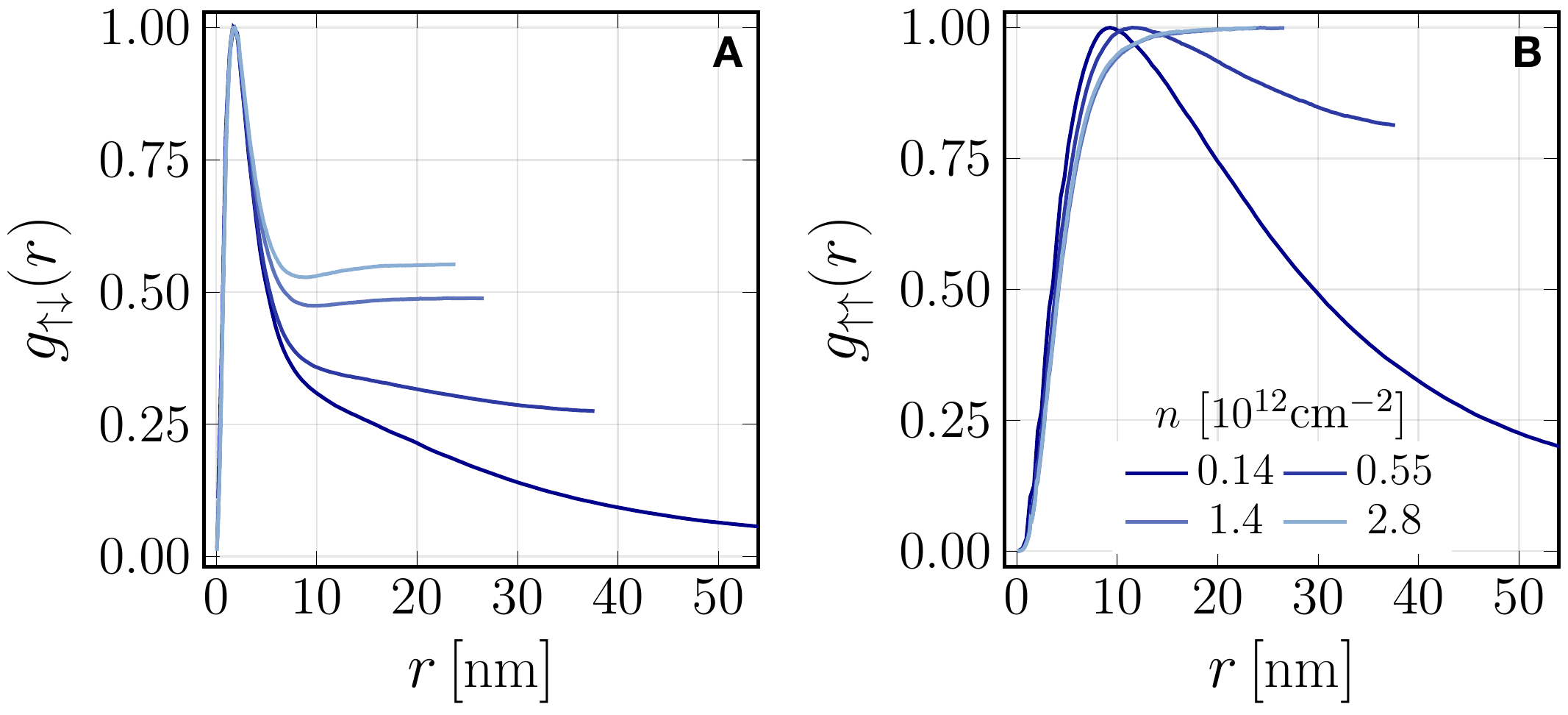}

    \caption{\textbf{Exciton spatial correlation functions in the droplet phase.} Panels (\textbf{A}) and (\textbf{B}) show $g_{\uparrow\downarrow}(r)$ and $g_{\uparrow\uparrow}(r)$, respectively, for several exciton densities in the droplet region, as indicated in the legend of (\textbf{B}). Calculations were performed with $N=32$ and $T=1.5$~K.
    }
    \label{fig:droplet_grs}
\end{figure}

Next, we examine dipolar correlations within the droplet state. In Fig.~\ref{fig:droplet_grs}, we compare the spatial correlations between excitons with antiparallel and parallel dipole moments across several densities within the droplet state. We find that antiparallel dipolar correlations occur at significantly shorter interexciton distances and persist to higher densities compared to the parallel case. These results indicate antiparallel dipolar ordering between neighboring excitons, further emphasizing antiparallel dipolar attraction as the stabilizing mechanism of the droplet.

\begin{figure}[ht!]
    \centering
    \includegraphics[width=0.95\textwidth]{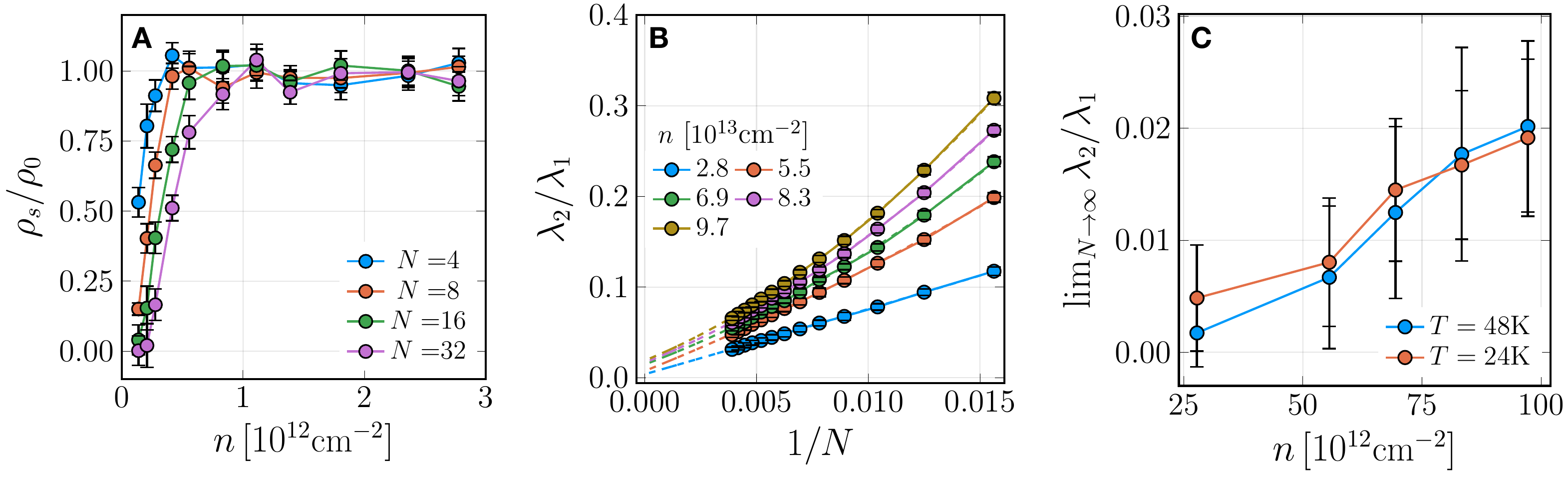}

    \caption{\textbf{Characteristics of the partially fragmented superfluid phase.} (\textbf{A}) Density dependence of the superfluid stiffness for increasing system sizes, computed at temperature $T=1.5$~K. (\textbf{B}) System size dependence of the eigenvalue ratio at high exciton densities for $T=24$~K. Dashed lines correspond to the fitting function $y=a/x^2 + b/x +c$. \textbf{(C)} Temperature convergence of the extrapolated $N\to\infty$ values of $\lambda_2/\lambda_1$ as a function of exciton density.}
    \label{fig:frag_SF_conv}
\end{figure}

As the exciton density increases, the droplet melts because of increased dipolar repulsion. This melting transition is manifested in the evolution of the superfluid stiffness as a function of density, shown in Fig.~\ref{fig:frag_SF_conv}A. At low densities, $\rho_s$ systematically decreases with increasing system size, implying a vanishing value in the thermodynamic limit. We note that, within our simulation framework, $\rho_s$ is evaluated from the winding number distribution, which corresponds to the system's response to global translations. For spatially homogeneous phases, a vanishing winding number implies a normal liquid state. In contrast, the absence of winding in the droplet phase originates from its spatial inhomogeneity. Thus, this observation provides an additional signature of the droplet. Identifying possible superfluid features within the droplet requires analyzing the local response to rotations, which we leave for future studies.

At higher densities, $\rho_s$ approaches its expected zero-temperature value at all system sizes, marking the melting of the droplet into a high-density superfluid state. As discussed in the main text, the resulting high-density condensate exhibits a complex internal structure related to the dipole moment orientation degree of freedom. To characterize this structure, we compute the ratio between the eigenvalues of the dipole-orientation-resolved superfluid fraction tensor, $\lambda_2/\lambda_1$. In Fig.~\ref{fig:frag_SF_conv}B, we present a finite-size analysis of the $\lambda_2/\lambda_1$ ratio for exciton densities within the superfluid region. We observe a monotonic decay of the ratio with increasing system size, indicating the suppression of dipolar fluctuations. Nevertheless, we find that $\lambda_2/\lambda_1$ remains finite for all measured system sizes, suggesting a partial fragmentation that is stable even for a considerably large system with $N=256$ particles. To extrapolate to the thermodynamic limit $N\to\infty$, we employ a scaling function $y=\frac{a}{x^2} + \frac{b}{x} + c$, which yields finite and density-increasing values of $\lambda_2/\lambda_1$, as shown in Figs.~\ref{fig:frag_SF_conv}C and in Fig.~4F of the main text.

To ensure that our results for the partially fragmented superfluid reflect ground-state properties, the simulation temperatures were systematically scaled according to the BKT transition temperature, $T_{\text{BKT}} \propto n/m_X$. Using the critical temperature of the quadrupolar superfluid transition ($T_{\text{BKT}} \approx 17$~K at $n = 1.4 \cdot 10^{12}\text{ cm}^{-2}$) as a baseline, the substantially higher densities of the partially fragmented regime ($n \sim 10^{13} - 10^{14}\text{ cm}^{-2}$) yield characteristic BKT scales ranging from $T_{\text{BKT}} \sim 170$~K to $1700$~K. Consequently, our simulations performed at $T=24$~K and $T=48$~K can be considered representative of the asymptotic ground-state limit ($T \ll T_{\text{BKT}}$). To explicitly verify this, we repeat the calculation of the extrapolated values at $T=48$~K and $T=24$~K. As presented in Fig.~\ref{fig:frag_SF_conv}C, the results agree within the error bars, numerically confirming the low-temperature convergence of our data.

\begin{figure}[t!]
    \centering
    \includegraphics[width=0.65\textwidth]{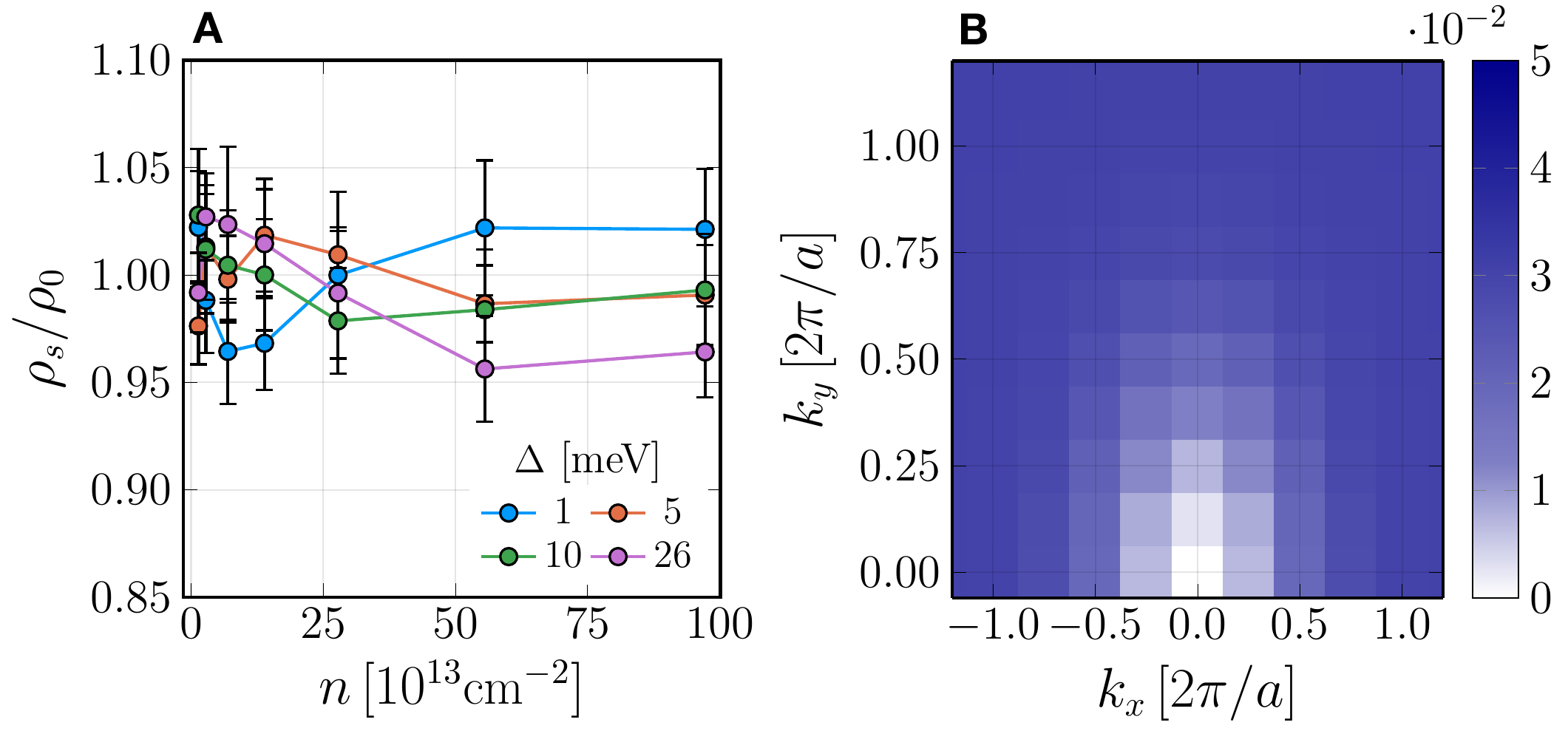}
    \caption{\textbf{Absence of a crystal phase at experimental exciton mass.} (\textbf{A}) The superfluid stiffness as a function of exciton density and the dipolar-quadrupolar gap, as denoted in the legend. The results are presented for $N=32$ and temperatures below the BKT transition for each density value, ranging between $24$~K$\leq T \leq 192$~K. (\textbf{B}) A two-dimensional momentum map of the structure factor, shown for $n=10^{15}$~cm$^{-2}$ and $\Delta=1$~meV. Similar momentum maps were obtained for all exciton densities and $\Delta$ values shown in (\textbf{A}).}
    \label{fig:noLatt}
\end{figure}

To conclude this section, we note that no crystalline phase was detected at the experimentally relevant exciton mass, $m_X=m_0$. We demonstrate this result in Fig.~\ref{fig:noLatt} by presenting the superfluid stiffness and a two-dimensional momentum-space map of the total structure factor for dipolar flipping rates between $\Delta=1{-}26$~meV and exciton densities up to $n=10^{15}$~cm$^{-2}$. At all parameters, we observe finite superfluid stiffness accompanied by a homogeneous form of $S(\mathbf{k})$, indicative of the absence of spatial order. These results confirm that zero-point motion is significant at the considered exciton mass and prevents the formation of a crystal. Thus, we conclude that stabilizing an exciton crystal requires larger exciton masses, where quantum fluctuations are sufficiently suppressed.

\subsection*{Staggered dipolar crystal}

In this section, we further characterize the staggered dipolar crystal at high exciton mass. We begin by analyzing finite-size and finite-temperature effects for the crystal at exciton density $n=1.4\cdot 10^{14}$~cm$^{-2}$ and mass $m_X=5m_0$, as discussed in the main text.

\begin{figure}[ht!]
    \centering
    \includegraphics[width=0.65\textwidth]{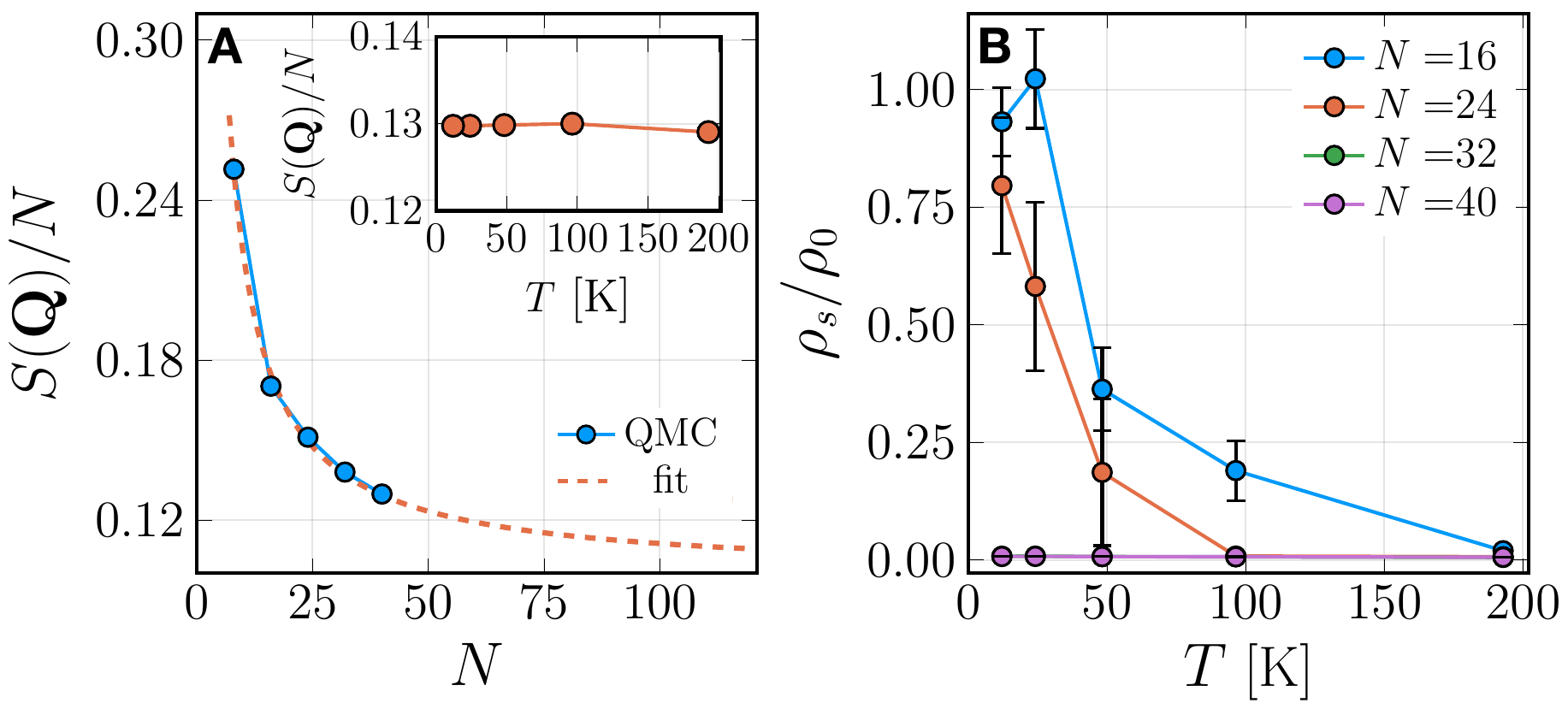}
    \caption{\textbf{Finite-size and finite-temperature analysis of the crystal phase.} (\textbf{A}) The structure factor evaluated at the Bragg vector $\mathbf{Q}$ as a function of system size, computed at temperature $T=12$~K. Blue dots denote QMC results. The red dashed line corresponds to the fitting function $y=a/x + b$, from which we obtain the $N\to \infty$ value of the Bragg peak, $b=0.099 \pm 0.002$. The inset shows the temperature dependence of the Bragg peak at $N=40$. (\textbf{B}) The superfluid stiffness as a function of temperature and system size. In both panels, the exciton density is set to $n=1.4\cdot 10^{14}$~cm$^{-2}$.}
     \label{fig:SQ_rho_conv}
\end{figure}

In Fig.~\ref{fig:SQ_rho_conv}A, we present the total structure factor, evaluated at the observed Bragg wavevector, as a function of system size. $S(\mathbf{Q})$ systematically converges with system size, as demonstrated by the comparison with the fitting function $y=a/x+b$. The fit yields a finite value of $b=0.099\pm0.002$, indicating the presence of crystalline order in the thermodynamic limit. Furthermore, we examine the temperature dependence of $S(\mathbf{Q})$ for a system containing $N=40$ particles and observe negligible variations, as displayed in the inset of Fig.~\ref{fig:SQ_rho_conv}A. This behavior confirms the ground-state convergence of our results.

Further evidence for the crystalline phase arises from the absence of superfluidity in the thermodynamic limit, as reflected by the vanishing superfluid stiffness at large system sizes, presented in Fig.~\ref{fig:SQ_rho_conv}B. Based on the above analysis, we set $N=40$, for which residual finite-size effects are sufficiently small, and take $T=48$~K as our low-temperature limit. 

To further justify this result, we evaluate the melting transition of the staggered dipolar crystal using the anisotropic generalization of the KTHNY criterion \cite{Podolsky_melting_2026},
\begin{equation}
16\pi = \frac{Ya^2}{k_BT_m\sqrt{1+\zeta/4}} \,,
\end{equation}
where $Y$ is the two-dimensional Young's modulus, $\zeta$ is the anisotropy parameter and $a=1/\sqrt{n}$ is the lattice constant. 
To compute these coefficients, we use a harmonic model with a full lattice sum over pairwise interactions; this includes, in particular, the nearest-neighbor antiparallel bonds and the next-nearest-neighbor same-sublattice repulsive bonds required for mechanical stability. 
This yields a melting temperature of $T_m\approx 178$ K for the characteristic density of the crystal, $n=1.4\cdot 10^{14}\text{ cm}^{-2}$. However, this harmonic calculation does not fully capture the substantial zero-point motion observed in the staggered crystal (see Fig.~17C), which is expected to introduce anharmonic corrections to the phononic dynamics. Thus, the melting temperature may be modified. Nevertheless, the above estimate for $T_m$ corroborates the numerical results presented in Fig.~\ref{fig:SQ_rho_conv}, further establishing $T=48$~K as a representative ground-state temperature.

\begin{figure}[ht!]
    \centering
    \includegraphics[width=0.33\textwidth]{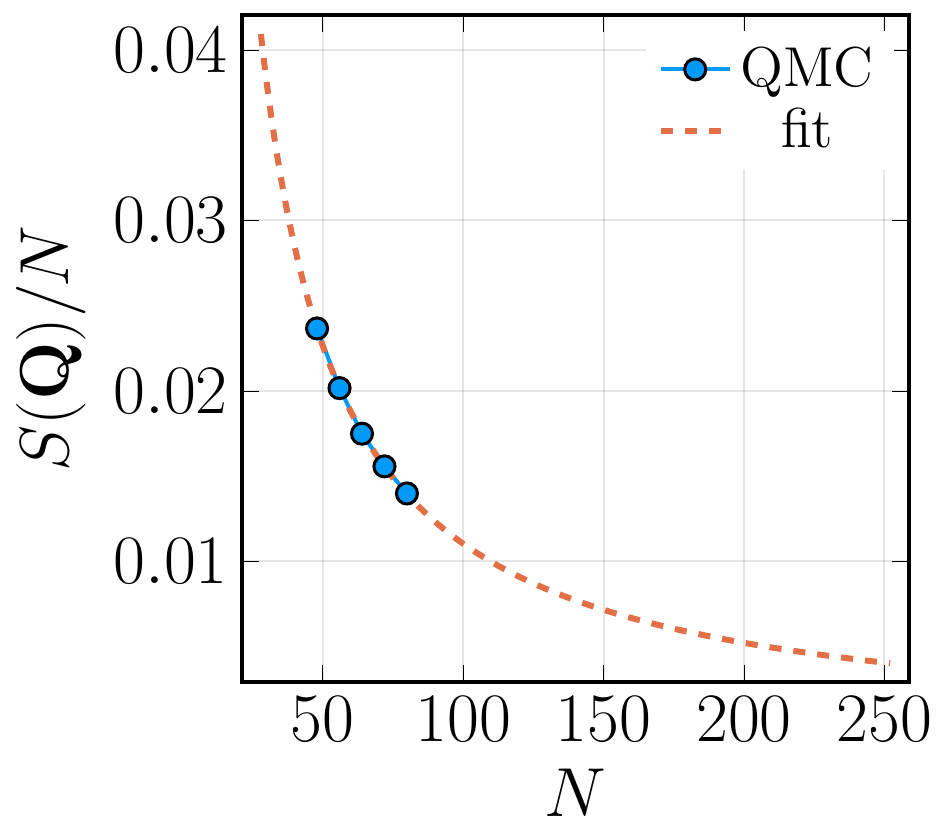}
    \caption{\textbf{Finite-size analysis of the crystal phase at finite electric field.} The structure factor evaluated at the Bragg vector $\mathbf{Q}$ as a function of system size, computed at exciton density $n=1.4\cdot 10^{14}$~cm$^{-2}$, electric field value $E_z=0.53$~V\,nm$^{-1}$, and temperature $T=48$~K. Blue dots mark the QMC results, and the red dashed line represents the fit to the data, $ y = a/x + b$. In the thermodynamic limit, the Bragg peak value is $b=(5\pm 1)\cdot 10^{-4}$, as obtained from the fitting function.}
    \label{fig:SQ_finite_Ez_conv}
\end{figure}

Next, we consider the effect of a finite out-of-plane electric field applied to the staggered crystal. In the main text, we show that $E_z$ drives a crystal-to-superfluid melting transition, characterized by a drop in $S(\mathbf{Q})$ correlated with a rise in $\rho_s$ with increasing $E_z$. The high $E_z$ superfluid phase is identified by the expected superfluid stiffness value $\rho_s=\rho_0$, as shown in Fig.~5G of the main text, and by a vanishing structure factor. To verify the latter, we test the convergence of our data with system size, setting a finite value of the electric field within the superfluid region, $E_z=0.53$~V\,nm$^{-1}$. The Bragg peak value displays a $\propto 1/N$ convergence, as evident from the agreement between the QMC data and the fitting function $y$ (defined above) in Fig.~\ref{fig:SQ_finite_Ez_conv}. The fit yields $b=(5\pm 1)\cdot 10^{-4}$, confirming the vanishing of the structure factor in the thermodynamic limit.

\begin{figure}[ht!]
    \centering
    \includegraphics[width=0.33\textwidth]{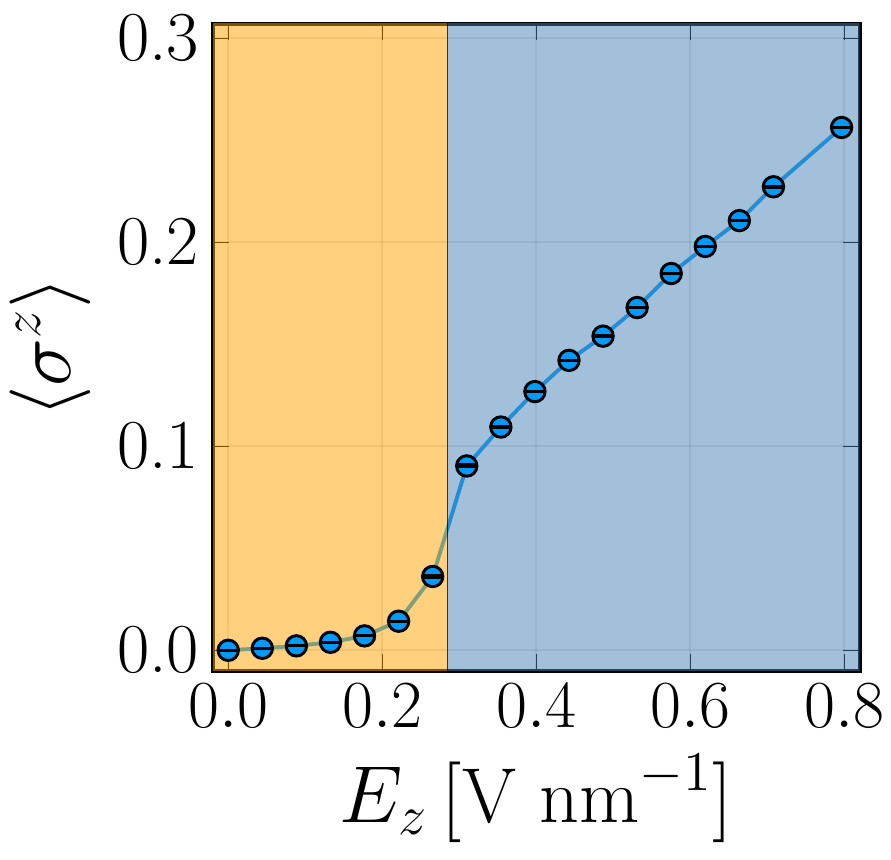}
    \caption{\textbf{Average exciton dipole moment across the crystal to superfluid melting transition.} The average dipole moment per exciton as a function of $E_z$, plotted for an exciton density $n=1.4\cdot 10^{14}$~cm$^{-2}$, system size $N=40$ and temperature $T=48$~K. The staggered crystal and superfluid regions are indicated by orange and blue backgrounds, respectively.}
    \label{fig:avg_d_crystal}
\end{figure}

To further substantiate the crystal to superfluid melting transition, we calculate the average dipole moment per exciton $\expval{\sigma_z}$ as a function of the electric field. The results are presented in Fig.~\ref{fig:avg_d_crystal}. We find that, at weak bias, $\expval{\sigma_z}\approx0$, confirming the stability of the staggered crystal phase. As $E_z$ increases, $\expval{\sigma_z}$ sharply increases around $E_z=0.3$~V\,nm$^{-1}$, coinciding with the onset of superfluidity, as depicted in Fig.~5G of the main text. At larger electric fields, $\expval{\sigma_z}$ grows linearly, indicating the progressive polarization of the excitons in the superfluid phase.

\begin{figure}[ht!]
    \centering
    \includegraphics[width=0.9\textwidth]{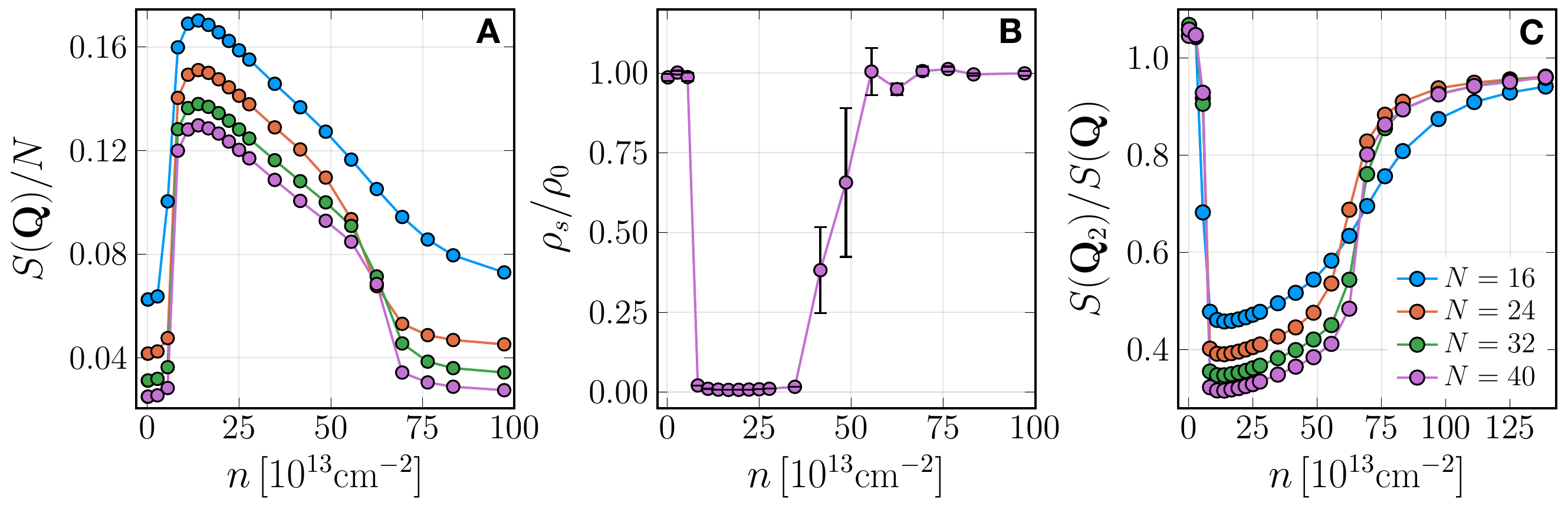}

    \caption{\textbf{Melting of the staggered crystal at low and high densities.} Dependence of (\textbf{A}) the Bragg peak, (\textbf{B}) the superfluid stiffness, and (\textbf{C}) the ratio between the secondary and primary Bragg peaks of $S(\mathbf{k})$ on exciton density. Different colors in (\textbf{A}) and (\textbf{C}) correspond to increasing system sizes, as denoted in the legend of (\textbf{C}). Panel (\textbf{B}) is shown for $N=40$ and temperatures below the BKT transition at each density. 
    } 
    \label{fig:crystal_n_melting}
\end{figure}

We proceed by studying the stability of the staggered crystal as a function of exciton density. In Fig.~\ref{fig:crystal_n_melting}A, we compute the evolution of $S(\mathbf{Q})$ with exciton density and system size. A finite Bragg-peak value is observed in the density range $5.5\cdot10^{13}$~cm$^{-2}$$\leq n \leq5.5\cdot10^{14}$~cm$^{-2}$, directly indicating a crystal state. Beyond this density region, we observe a finite superfluid stiffness, as shown in Fig.~\ref{fig:crystal_n_melting}B, marking exciton condensation in both the low- and high-density limits. To understand this behavior, we note that the crystal is stabilized by attractive interactions between antiparallel dipoles, which are maximized at separation $r=1.67d$ (equivalent to $1$~nm for $d=0.6$~nm). At larger distances, the interaction decays as $1/r^3$. Consequently, in the dilute density limit, the kinetic energy term becomes dominant, melting the crystal into a superfluid phase. The high-density reentrance phenomenon is more surprising. As the exciton density increases, antiparallel attraction turns repulsive at short distances, eventually destabilizing the staggered crystal pattern and driving a second melting transition into a superfluid state. 

To assess the crystal's stability under quantum fluctuations, we quantify the strength of exciton zero-point motion in this phase. Specifically, we compare the amplitudes of the first and second Bragg peaks in $S(\mathbf{k})$, located at $\mathbf{Q}$ and $\mathbf{Q}_{2}$, respectively. In the absence of zero-point motion, $S(\mathbf{k})$ reduces to a sum of delta functions at the reciprocal lattice sites. Thus, the Bragg-peak ratio $S(\mathbf{Q}_2)/S(\mathbf{Q})$ equals unity. In the opposite limit of strong quantum fluctuations, higher-order Bragg peaks are expected to vanish, yielding $S(\mathbf{Q}_2)/S(\mathbf{Q})\to 0$ \cite{Gazit_2016}.

In Fig.~\ref{fig:crystal_n_melting}C, we display the ratio between the secondary and primary Bragg peaks as a function of exciton density for several system sizes. We find that, within the staggered crystal region, the ratio converges to a relatively small value ($<0.4$), indicating significant zero-point motion of the excitons. At lower and higher densities, the ratio increases with the melting of the crystal into superfluid phases.

\begin{figure}[ht!]
    \centering
    \includegraphics[width=1.\textwidth]{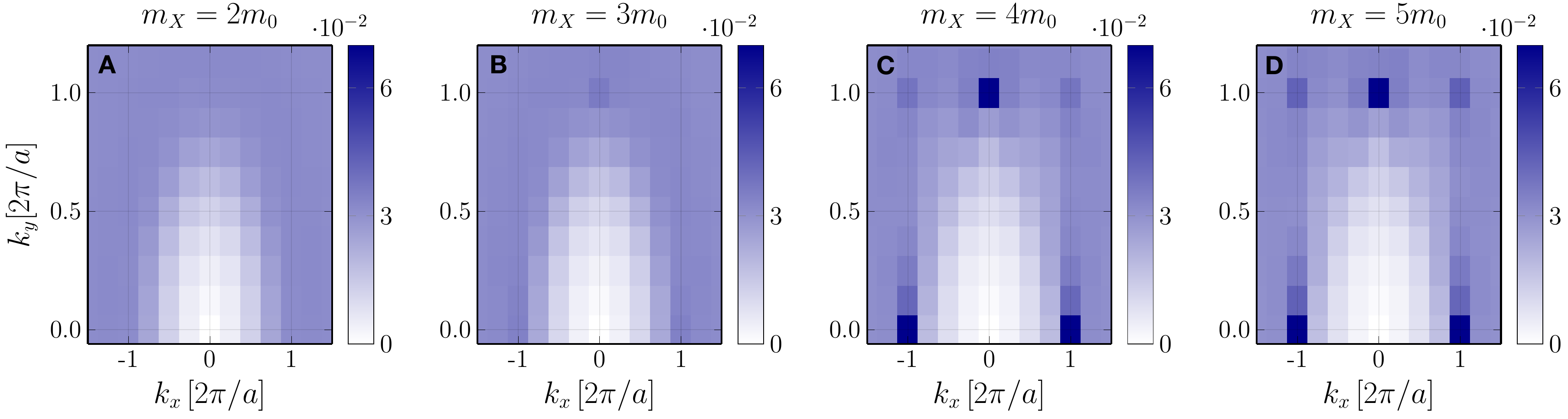}

    \caption{\textbf{Onset of a staggered crystal phase.} Two-dimensional momentum-space maps of the total structure factor $S(\mathbf{k})$ computed for exciton masses $m_X=2{-}5m_0$, as indicated in the title of each panel, and $N=32$ excitons. 
    }
    \label{fig:stg_crystal_onset}
\end{figure}

The strong quantum fluctuations require relatively large exciton masses to stabilize a staggered crystal phase. To characterize the onset of a crystal phase, we scan exciton masses above the experimental value $m_X=m_0$. To this end, we compute two-dimensional momentum-space maps of the total structure factor, as presented in Fig.~\ref{fig:stg_crystal_onset}. We find that $S(\mathbf{k})$ develops pronounced Bragg peaks at exciton masses $m_X \gtrsim 4m_0$. At lower masses, the momentum maps are radially uniform, indicating a liquid state.

\subsection*{Spatial correlation functions}

\begin{figure}[ht!]
    \centering
    \includegraphics[width=0.9\textwidth]{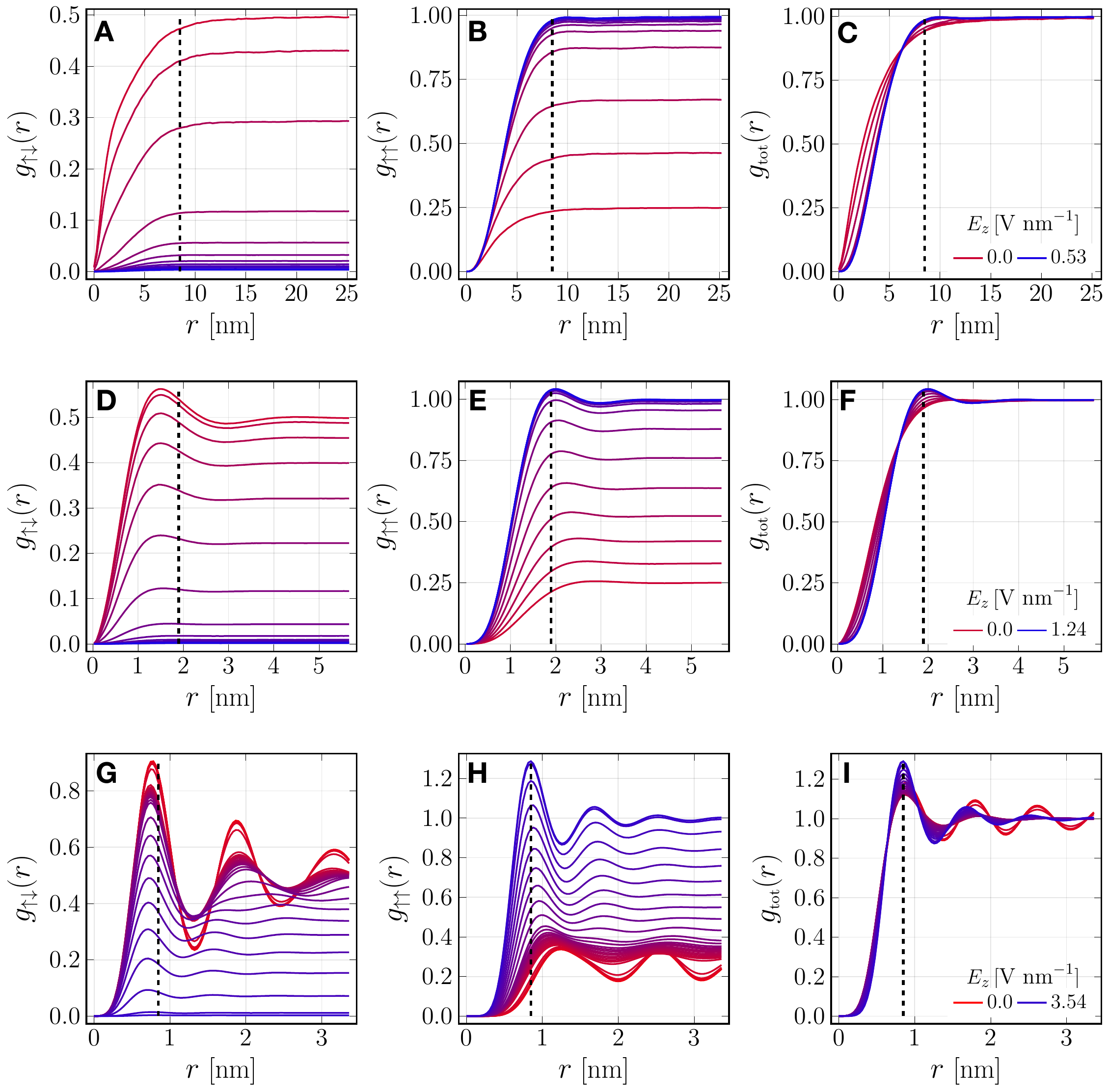}

  \caption{\textbf{Exciton spatial correlations in the low and high density liquid phase and in the staggered crystal.} Evolution of exciton spatial correlation functions with the out-of-plane electric field $E_z$. 
  Columns from left to right show antiparallel, parallel, and total spatial correlation functions. Rows correspond to different physical regimes: (\textbf{A})-(\textbf{C}) a dilute quadrupolar liquid ($n=1.4\times10^{12}$~cm$^{-2}$, $m_X=m_0$); (\textbf{D})-(\textbf{F}) a high-density liquid ($n=2.8\times10^{13}$~cm$^{-2}$, $m_X=m_0$); and (\textbf{G})-(\textbf{I}) a staggered dipolar crystal ($n=1.4\times10^{14}$~cm$^{-2}$, $m_X=5m_0$).
  }
  \label{fig:gr_phases}
\end{figure}

To gain insight into exciton spatial correlations across different regimes of our model, we examine their dependence on an applied out-of-plane electric field. For completeness, in addition to the dipole-orientation-resolved correlations, we compute the total exciton spatial correlation function, $g_{\text{tot}}(r)$, obtained by summing the contributions from all relative dipole orientations.

In Fig.~\ref{fig:gr_phases}, we present the dipole-moment-resolved and total spatial correlation functions as a function of $E_z$, for the low- and high-density liquid states, as well as the staggered dipolar exciton crystal. In the dilute density limit, excitons form a weakly interacting liquid. In this phase, interexciton correlations are suppressed, as reflected by the absence of correlation peaks in Figs.~\ref{fig:gr_phases} (A)-(C). At zero bias, parallel and antiparallel configurations are equally probable, consistent with the quadrupolar nature of the excitons in this case. As $E_z$ increases, the relative weight of $g_{\uparrow\uparrow}(r)$ grows on account of other dipolar configurations, as expected from the effect of dipole-orientation polarization.

At increased exciton densities, dipolar interactions give rise to spatial correlations between the excitons, as evident in the emergent correlation peaks in Fig.~\ref{fig:gr_phases} (D)-(F).
Notably, in the unbiased case, antiparallel and parallel dipolar correlations peak at separations smaller and larger than the mean interexciton distance, respectively. This result stems from the attractive (antiparallel) and repulsive (parallel) nature of dipolar interactions, underscoring their importance in the high-density limit. As $E_z$ increases, the primary peak of parallel correlations shifts towards shorter distances, until it coincides with the mean interexciton distance. This process reflects the gradual polarization of the exciton dipole moment by the external electric field.

These trends in the spatial correlation functions are more pronounced in the staggered dipolar crystal, as shown in Fig.~\ref{fig:gr_phases} (G)-(I). In this case, the primary peaks of $g_{\uparrow\downarrow}$ and $g_{\uparrow\uparrow}$ correspond to the nearest neighbor exciton with an opposite and parallel dipole moment, respectively. Therefore, the shift of the primary peak of $g_{\uparrow\uparrow}$ to smaller separations additionally signifies the electric-field-induced melting of the crystal. This transition is further evidenced by the gradual suppression of higher-order correlation peaks and the increasing dominance of parallel dipolar configurations.

\section{Additional exact diagonalization results \label{app:ED}}

\subsection*{Two excitons with antiparallel dipole moments}

In the main text, we have shown that the exciton droplet state is characterized by a density-independent interexciton separation. To provide additional insight into this interexciton distance, we consider the two-body limit of interacting excitons with fixed and opposite dipole moments. This model accounts for the dominant antiparallel dipolar correlations in the droplet by taking the limit of vanishing dipole moment fluctuations. 

\begin{figure}[ht!]
    \centering
    \includegraphics[width=0.34\textwidth]{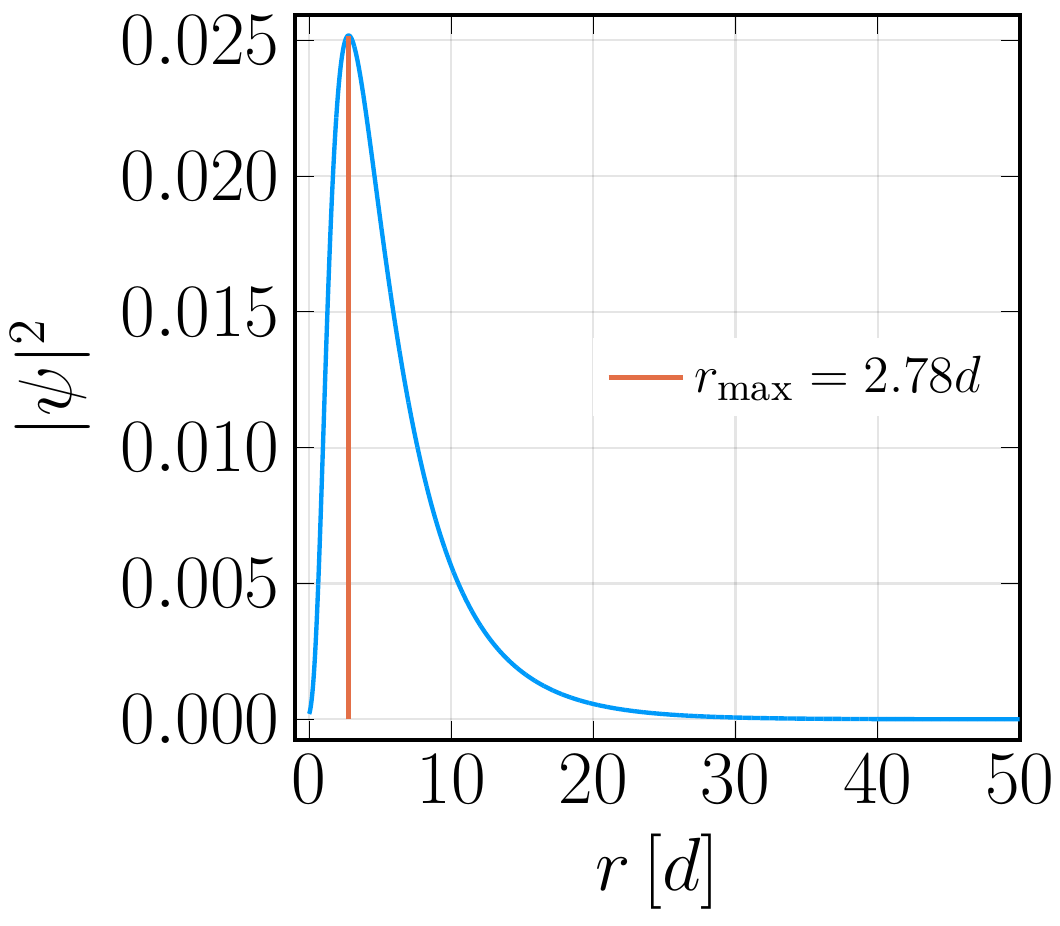}
    \vspace{-0.5mm}
    \caption{\textbf{Probability distribution for a pair of antiparallel dipoles.} The ground-state spatial probability distribution function of a two-body model consisting of dipolar bosons with fixed and opposite dipole moments. The distribution function peaks at $r_{\text{max}}=2.78d$. This result was obtained for a typical exciton mass of $m_X=m_0$.}
    \label{fig:opposite_2body}
\end{figure}

To this end, we solve the two-body model in Eq.~\ref{eqn:H_2bd} in the limit of $\Delta\to 0$. The solution reveals the existence of a bound state between two oppositely oriented dipoles, as evidenced by the spatial probability distribution function shown in Fig.~\ref{fig:opposite_2body}. The distribution function exhibits a maximum at $r=2.78d$, corresponding to $1.67$~nm for a dipole length of $d=0.6$~nm. This value is in good agreement with the mean interexciton separation in the droplet obtained from QMC simulations, thus confirming the importance of antiparallel dipolar attraction in stabilizing this phase.

\subsection*{Static two-body problem}

Here, we provide additional details on the simplified two-body model discussed in the main text. Our model comprises two static dipolar excitons, with allowed dipole moment flips, residing at a fixed relative in-plane distance $\Delta r$. The excitons interact via the dipolar potential defined in Eq.~2 and are subjected to an externally applied out-of-plane electric field. The model Hamiltonian reads

\begin{equation} \label{eqn:H_stat}
    H_2 =  - \Delta \left(\hat{\sigma}_1^x +\hat{\sigma}_2^x \right) - edE_z \left(\hat{\sigma}_1^z + \hat{\sigma}_2^z\right) +  U^d_{\sigma^z_1,\sigma^z_2}(\Delta r) \,.
\end{equation}
The above Pauli operators are defined in the single-particle dipolar basis. The exciton chemical potential is given by the energy difference between systems with $N=2$ and $N=1$ excitons, that is, $\mu_X = E^{gs}_2 - E^{gs}_1$, where $E^{gs}_{1}$ and $E^{gs}_{2}$ are the ground-state energies of the single- and two-exciton systems, respectively. We obtain $E^{gs}_2$ by diagonalizing Eq.~\ref{eqn:H_stat}, and obtain $E^{gs}_1$ by solving the following single-particle Hamiltonian:
\begin{equation} \label{eqn:H_1bd}
    H_1 =  - \Delta \hat{\sigma}^x  - edE_z \hat{\sigma}^z  \,.
\end{equation}

\begin{figure}[h!]
    \centering
    \includegraphics[width=0.33\textwidth]{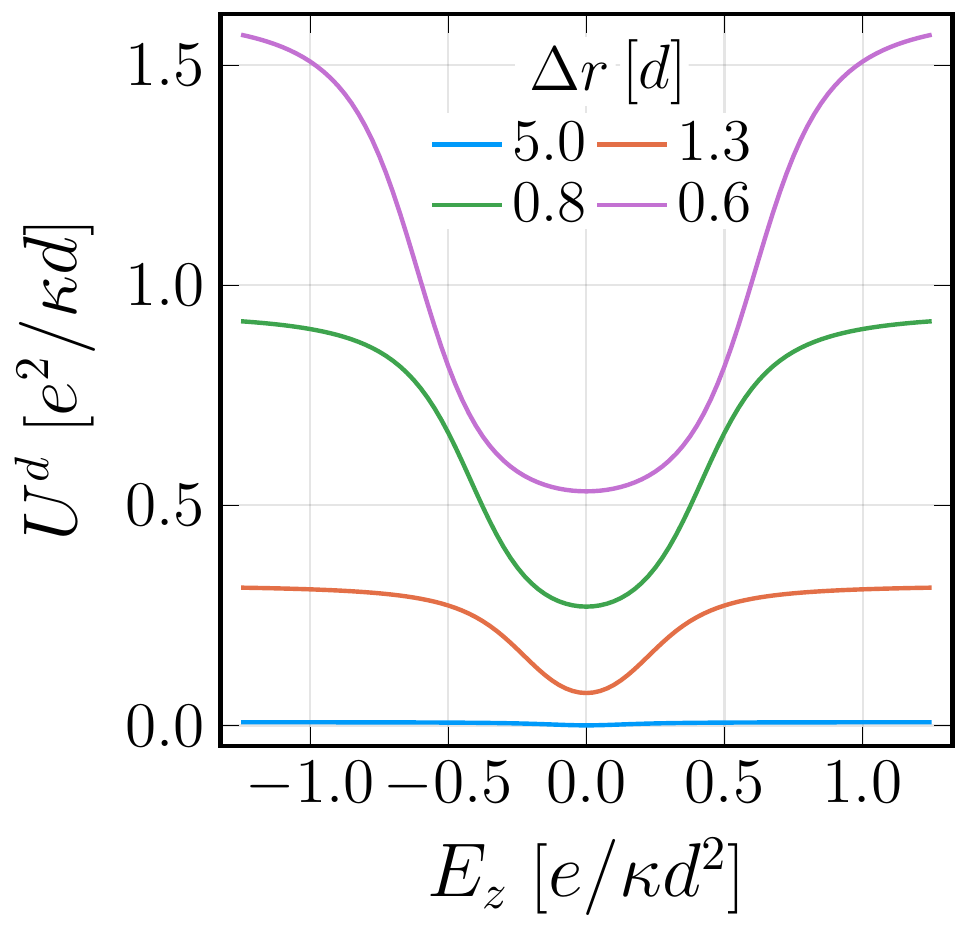}
    \caption{\textbf{Static two-body pair interaction energy.} The ground-state interparticle interaction for the model described in Eq.~\ref{eqn:H_stat}, shown as a function of $E_z$ for several interparticle separations $\Delta r$. This dipole-flipping rate is fixed at $\Delta=0.2\varepsilon$. 
    }
  \label{fig:U_stat_ED}
\end{figure}

In Fig.~3C of the main text, we show the resulting dependence of $\mu_X$ on the out-of-plane electric field for $\Delta=0.2\varepsilon$ and several interexciton separations $\Delta r$. To elucidate the transition in $\mu_X(E_z)$ from a redshift to a blueshift with decreasing interexciton separation, we compute the interparticle interaction energy in the ground state of Eq.~\ref{eqn:H_stat} as a function of $E_z$. The results are presented in Fig.~\ref{fig:U_stat_ED}. We find that the interparticle interaction increases with $|E_z|$, pointing to enhanced dipolar repulsion as the dipoles gradually polarize. This trend becomes more pronounced at smaller $\Delta r$, consistent with the increasing strength of the dipolar interaction. On the other hand, at sufficiently large distances, interexciton interactions are negligible, and the system is governed by the dipole-flip term, leading to quadrupolar behavior.

These results directly relate to the exciton chemical potential. The computed ground-state interaction energy corresponds to the energy penalty of adding a second exciton to a system described by Eq.~\ref{eqn:H_1bd}. Hence, the large dipolar repulsion at strong $E_z$ manifests in the blueshift of $\mu_X(E_z)$, which is further enhanced at small interexciton separations.

\twocolumngrid
\bibliography{bib_arxiv}

@Article{Slobodkin_2020,
  title = {Quantum Phase Transitions of Trilayer Excitons in Atomically Thin Heterostructures},
  author = {Slobodkin, Yevgeny and Mazuz-Harpaz, Yotam and Refaely-Abramson, Sivan and Gazit, Snir and Steinberg, Hadar and Rapaport, Ronen},
  journal = {Phys. Rev. Lett.},
  volume = {125},
  issue = {25},
  pages = {255301},
  numpages = {7},
  year = {2020},
  month = {Dec},
  publisher = {American Physical Society},
}

@Article{Zhang_2024,
	author = {Zhang, Zhe and Wang, Shudong},
	journal = {The Journal of Physical Chemistry Letters},
	number = {40},
	pages = {10104-10110},
	title = {Diverse Excitonic Phenomena in Asymmetric Trilayer Transition Metal Dichalcogenide Heterostructures},
	volume = {15},
	year = {2024},
}

@Article{Deilmann_2024,
	author = {Deilmann, Thorsten and Sommer Thygesen, Kristian},
	journal = {2D Materials},
	month = {jun},
	number = {3},
	pages = {035032},
	publisher = {IOP Publishing},
	title = {Quadrupolar and dipolar excitons in symmetric trilayer heterostructures: insights from first principles theory},
	volume = {11},
	year = {2024},
}

@Article{Baym_2006,
  title = {Fragmentation of {B}ose-{E}instein condensates},
  author = {Mueller, Erich J. and Ho, Tin-Lun and Ueda, Masahito and Baym, Gordon},
  journal = {Phys. Rev. A},
  volume = {74},
  issue = {3},
  pages = {033612},
  numpages = {17},
  year = {2006},
  month = {Sep},
  publisher = {American Physical Society},
}

@Article{Nelson_1977,
  title = {Universal Jump in the Superfluid Density of Two-Dimensional Superfluids},
  author = {Nelson, David R. and Kosterlitz, J. M.},
  journal = {Phys. Rev. Lett.},
  volume = {39},
  issue = {19},
  pages = {1201--1205},
  numpages = {0},
  year = {1977},
  publisher = {American Physical Society},
}

@Article{Nelson_2014,
  title = {Quantum hexatic order in two-dimensional dipolar and charged fluids},
  author = {Bruun, Georg M. and Nelson, David R.},
  journal = {Phys. Rev. B},
  volume = {89},
  issue = {9},
  pages = {094112},
  numpages = {10},
  year = {2014},
  publisher = {American Physical Society},
}

@Article{Penrose_1956,
  title = {Bose-{E}instein Condensation and Liquid Helium},
  author = {Penrose, Oliver and Onsager, Lars},
  journal = {Phys. Rev.},
  volume = {104},
  issue = {3},
  pages = {576--584},
  numpages = {0},
  year = {1956},
  month = {Nov},
  publisher = {American Physical Society},
}

@Article{Astrakharchik_2007,
  title = {Quantum Phase Transition in a Two-Dimensional System of Dipoles},
  author = {Astrakharchik, G. E. and Boronat, J. and Kurbakov, I. L. and Lozovik, Yu. E.},
  journal = {Phys. Rev. Lett.},
  volume = {98},
  issue = {6},
  pages = {060405},
  numpages = {4},
  year = {2007},
  publisher = {American Physical Society},
}

@Article{Buchler_2007,
  title = {Strongly Correlated 2D Quantum Phases with Cold Polar Molecules: Controlling the Shape of the Interaction Potential},
  author = {B\"uchler, H. P. and Demler, E. and Lukin, M. and Micheli, A. and Prokof'ev, N. and Pupillo, G. and Zoller, P.},
  journal = {Phys. Rev. Lett.},
  volume = {98},
  issue = {6},
  pages = {060404},
  numpages = {4},
  year = {2007},
  month = {Feb},
  publisher = {American Physical Society},
}

@Article{Laikhtman_2009,
  title = {Exciton correlations in coupled quantum wells and their luminescence blue shift},
  author = {Laikhtman, B. and Rapaport, R.},
  journal = {Phys. Rev. B},
  volume = {80},
  issue = {19},
  pages = {195313},
  numpages = {12},
  year = {2009},
  month = {Nov},
  publisher = {American Physical Society},
}

@Article{Cohen_2016,
	author = {Cohen, Kobi and Shilo, Yehiel and West, Ken and Pfeiffer, Loren and Rapaport, Ronen},
	isbn = {1530-6984},
	journal = {Nano Letters},
	journal1 = {Nano Lett.},
	month = {06},
	number = {6},
	pages = {3726--3731},
	publisher = {American Chemical Society},
	title = {Dark High Density Dipolar Liquid of Excitons},
	volume = {16},
	year = {2016},
}

@Article {Mazuz_2019,
	author = {Mazuz-Harpaz, Yotam and Cohen, Kobi and Leveson, Michael and West, Ken and Pfeiffer, Loren and Khodas, Maxim and Rapaport, Ronen},
	title = {Dynamical formation of a strongly correlated dark condensate of dipolar excitons},
	volume = {116},
	number = {37},
	pages = {18328--18333},
	year = {2019},
	publisher = {National Academy of Sciences},
	issn = {0027-8424},
	journal = {Proceedings of the National Academy of Sciences}
}

@Article{Shilo_2013,
	author = {Shilo, Yehiel and Cohen, Kobi and Laikhtman, Boris and West, Ken and Pfeiffer, Loren and Rapaport, Ronen},
	isbn = {2041-1723},
	journal = {Nature Communications},
	number = {1},
	pages = {2335},
	title = {Particle correlations and evidence for dark state condensation in a cold dipolar exciton fluid},
	volume = {4},
	year = {2013},
}

@Article{Combescot_dark,
  title = {Bose-{E}instein Condensation in Semiconductors: The Key Role of Dark Excitons},
  author = {Combescot, Monique and Betbeder-Matibet, Odile and Combescot, Roland},
  journal = {Phys. Rev. Lett.},
  volume = {99},
  issue = {17},
  pages = {176403},
  numpages = {4},
  year = {2007},
  month = {Oct},
  publisher = {American Physical Society},
}

@Article{Combescot_gray,
  title = {``{G}ray'' {BCS} Condensate of Excitons and Internal Josephson Effect},
  author = {Combescot, Roland and Combescot, Monique},
  journal = {Phys. Rev. Lett.},
  volume = {109},
  issue = {2},
  pages = {026401},
  numpages = {5},
  year = {2012},
  publisher = {American Physical Society},
}

@Article{Dubin_2017,
  title = {Quantized Vortices and Four-Component Superfluidity of Semiconductor Excitons},
  author = {Anankine, Romain and Beian, Mussie and Dang, Suzanne and Alloing, Mathieu and Cambril, Edmond and Merghem, Kamel and Carbonell, Carmen Gomez and Lema\^{\i}tre, Aristide and Dubin, Fran{\c c}ois},
  journal = {Phys. Rev. Lett.},
  volume = {118},
  issue = {12},
  pages = {127402},
  numpages = {5},
  year = {2017},
  month = {Mar},
  publisher = {American Physical Society},
}

@Article{ButovCoherence2012,
	author = {High, A. A. and Leonard, J. R. and Hammack, A. T. and Fogler, M. M. and Butov, L. V. and Kavokin, A. V. and Campman, K. L. and Gossard, A. C.},
	isbn = {1476-4687},
	journal = {Nature},
	number = {7391},
	pages = {584--588},
	title = {Spontaneous coherence in a cold exciton gas},
	volume = {483},
	year = {2012},
}

@Article{Misra_2018,
  title = {Experimental Study of the Exciton Gas-Liquid Transition in Coupled Quantum Wells},
  author = {Misra, Subhradeep and Stern, Michael and Joshua, Arjun and Umansky, Vladimir and Bar-Joseph, Israel},
  journal = {Phys. Rev. Lett.},
  volume = {120},
  issue = {4},
  pages = {047402},
  numpages = {5},
  year = {2018},
  month = {Jan},
  publisher = {American Physical Society},
}

@Article{Misra_2022,
	author = {Subhradeep Misra and Michael Stern and Vladimir Umansky and Israel Bar-Joseph},
	journal = {Proceedings of the National Academy of Sciences},
	number = {32},
	pages = {e2203531119},
	title = {The Role of Spin-Flip Collisions in a Dark-Exciton Condensate},
	volume = {119},
	year = {2022},
}

@Article{Cutshall_2025,
	author = {Jacob Cutshall and Fateme Mahdikhany and Anna Roche and Daniel N. Shanks and Michael R. Koehler and David G. Mandrus and Takashi Taniguchi and Kenji Watanabe and Qizhong Zhu and Brian J. LeRoy and John R. Schaibley},
	journal = {Science Advances},
	number = {1},
	pages = {eadr1772},
	title = {Imaging interlayer exciton superfluidity in a 2D semiconductor heterostructure},
	volume = {11},
	year = {2025},
}

@Article{Yu_2023,
author = {Yu, Leo and Pistunova, Kateryna and Hu, Jenny and Watanabe, Kenji and Taniguchi, Takashi and Heinz, Tony F.},
da = {2023/12/01},
id = {Yu2023},
isbn = {1476-4660},
journal = {Nature Materials},
number = {12},
pages = {1485--1491},
title = {Observation of quadrupolar and dipolar excitons in a semiconductor heterotrilayer},
volume = {22},
year = {2023},
}

@Article{Li_2023,
author = {Li, Weijie and Hadjri, Zach and Devenica, Luka M. and Zhang, Jin and Liu, Song and Hone, James and Watanabe, Kenji and Taniguchi, Takashi and Rubio, Angel and Srivastava, Ajit},
id = {Li2023},
isbn = {1476-4660},
journal = {Nature Materials},
number = {12},
pages = {1478--1484},
title = {Quadrupolar--dipolar excitonic transition in a tunnel-coupled van der {W}aals heterotrilayer},
volume = {22},
year = {2023},
}

@Article{Xie_2023,
  title = {Bright and Dark Quadrupolar Excitons in the {W}{S}e$_{2}$/{M}o{S}e$_{2}$/{W}{S}e$_{2}$ Heterotrilayer},
  author = {Xie, Yongzhi and Gao, Yuchen and Chen, Fengyu and Wang, Yunkun and Mao, Jun and Liu, Qinyun and Chu, Saisai and Yang, Hong and Ye, Yu and Gong, Qihuang and Feng, Ji and Gao, Yunan},
  journal = {Phys. Rev. Lett.},
  volume = {131},
  issue = {18},
  pages = {186901},
  numpages = {6},
  year = {2023},
  month = {Oct},
  publisher = {American Physical Society},
}

@Article{Lian_2023,
author = {Lian, Zhen and Chen, Dongxue and Ma, Lei and Meng, Yuze and Su, Ying and Yan, Li and Huang, Xiong and Wu, Qiran and Chen, Xinyue and Blei, Mark and Taniguchi, Takashi and Watanabe, Kenji and Tongay, Sefaattin and Zhang, Chuanwei and Cui, Yong-Tao and Shi, Su-Fei},
isbn = {2041-1723},
journal = {Nature Communications},
number = {1},
pages = {4604},
title = {Quadrupolar excitons and hybridized interlayer {M}ott insulator in a trilayer moir{\'e} superlattice},
volume = {14},
year = {2023},
}

@Article{Xie_2024,
  title = {Transition between quadrupole and staggered dipole interlayer excitons in {W}{S}e$_{2}$/{M}o{S}e$_{2}$/{W}{S}e$_{2}$ heterotrilayers},
  author = {Xie, Yongzhi and Chen, Fengyu and Gao, Yuchen and Wang, Yunkun and Mao, Jun and Liu, Qinyun and Chu, Saisai and Yang, Hong and Ye, Yu and Gong, Qihuang and Feng, Ji and Gao, Yunan},
  journal = {Phys. Rev. B},
  volume = {110},
  issue = {20},
  pages = {L201402},
  numpages = {6},
  year = {2024},
  month = {Nov},
  publisher = {American Physical Society},
}

@Article{Bai_2023,
	author = {Bai, Yusong and Li, Yiliu and Liu, Song and Guo, Yinjie and Pack, Jordan and Wang, Jue and Dean, Cory R. and Hone, James and Zhu, Xiaoyang},
	isbn = {1530-6984},
	journal = {Nano Letters},
	journal1 = {Nano Lett.},
	m3 = {doi: 10.1021/acs.nanolett.3c03453},
	month = {12},
	number = {24},
	pages = {11621--11629},
	publisher = {American Chemical Society},
	title = {Evidence for Exciton Crystals in a 2D Semiconductor Heterotrilayer},
	volume = {23},
	year = {2023},
}

@Article{Meng_2025,
	author = {Meng, Yuze and Ma, Lei and Yan, Li and Khalifa, Ahmed and Chen, Dongxue and Zhang, Shuai and Banerjee, Rounak and Taniguchi, Takashi and Watanabe, Kenji and Tongay, Seth Ariel and Hunt, Benjamin and Lin, Shi-Zeng and Yao, Wang and Cui, Yong-Tao and Chatterjee, Shubhayu and Shi, Su-Fei},
	isbn = {1749-4893},
	journal = {Nature Photonics},
	number = {11},
	pages = {1219--1224},
	title = {Strong-interaction-driven quadrupolar-to-dipolar exciton transitions in a trilayer moir{\'e} superlattice},
	volume = {19},
	year = {2025},
}

@Article{Guo_2022,
	author = {Hongli Guo and Xu Zhang and Gang Lu},
	journal = {Science Advances},
	number = {40},
	pages = {eabp9757},
	title = {Tuning moir{\'e} excitons in Janus heterobilayers for high-temperature {B}ose-{E}instein condensation},
	volume = {8},
	year = {2022},
}

@Article{Brem_2020,
	author = {Brem, Samuel and Linder{\"a}lv, Christopher and Erhart, Paul and Malic, Ermin},
	isbn = {1530-6984},
	journal = {Nano Letters},
	m3 = {doi: 10.1021/acs.nanolett.0c03019},
	month = {12},
	number = {12},
	pages = {8534--8540},
	publisher = {American Chemical Society},
	title = {Tunable Phases of {M}oir{\'e} Excitons in van der {W}aals Heterostructures},
	volume = {20},
	year = {2020},
}

@Article{Hafezi_2024,
  title = {Long-Lived Topological Flatband Excitons in Semiconductor {M}oir\'e Heterostructures: A Bosonic {K}ane-{M}ele Model Platform},
  author = {Xie, Ming and Hafezi, Mohammad and Das Sarma, Sankar},
  journal = {Phys. Rev. Lett.},
  volume = {133},
  issue = {13},
  pages = {136403},
  numpages = {7},
  year = {2024},
  month = {Sep},
  publisher = {American Physical Society},
}

@Article{Malic_2020,
	author = {Brem, Samuel and Lin, Kai-Qiang and Gillen, Roland and Bauer, Jonas M. and Maultzsch, Janina and Lupton, John M. and Malic, Ermin},
	issue = {20},
	journal = {Nanoscale},
	pages = {11088-11094},
	publisher = {The Royal Society of Chemistry},
	title = {Hybridized intervalley moir{\'e} excitons and flat bands in twisted {W}{S}e$_2$ bilayers},
	volume = {12},
	year = {2020},
}

@Article{ZhengTrilayerMoire2023,
	author = {Zheng, Haihong and Wu, Biao and Li, Shaofei and Ding, Junnan and He, Jun and Liu, Zongwen and Wang, Chang-Tian and Wang, Jian-Tao and Pan, Anlian and Liu, Yanping},
	isbn = {2047-7538},
	journal = {Light: Science \& Applications},
	number = {1},
	pages = {117},
	title = {Localization-enhanced moir{\'e} exciton in twisted transition metal dichalcogenide heterotrilayer superlattices},
	volume = {12},
	year = {2023},
}

@Article{ChenBiTriMoire2022,
	author = {Chen, Dongxue and Lian, Zhen and Huang, Xiong and Su, Ying and Rashetnia, Mina and Yan, Li and Blei, Mark and Taniguchi, Takashi and Watanabe, Kenji and Tongay, Sefaattin and Wang, Zenghui and Zhang, Chuanwei and Cui, Yong-Tao and Shi, Su-Fei},
	id = {Chen2022},
	isbn = {2041-1723},
	journal = {Nature Communications},
	number = {1},
	pages = {4810},
	title = {Tuning moir{\'e} excitons and correlated electronic states through layer degree of freedom},
	volume = {13},
	year = {2022},
}

@Article{Fogler_2018,
	author = {Calman, E. V. and Fogler, M. M. and Butov, L. V. and Hu, S. and Mishchenko, A. and Geim, A. K.},
	id = {Calman2018},
	isbn = {2041-1723},
	journal = {Nature Communications},
	number = {1},
	pages = {1895},
	title = {Indirect excitons in van der Waals heterostructures at room temperature},
	volume = {9},
	year = {2018},
}

@Article{WatanabeTwist2021,
  title = {Twist Angle-Dependent Interlayer Exciton Lifetimes in van der {W}aals Heterostructures},
  author = {Choi, Junho and Florian, Matthias and Steinhoff, Alexander and Erben, Daniel and Tran, Kha and Kim, Dong Seob and Sun, Liuyang and Quan, Jiamin and Claassen, Robert and Majumder, Somak and Hollingsworth, Jennifer A. and Taniguchi, Takashi and Watanabe, Kenji and Ueno, Keiji and Singh, Akshay and Moody, Galan and Jahnke, Frank and Li, Xiaoqin},
  journal = {Phys. Rev. Lett.},
  volume = {126},
  issue = {4},
  pages = {047401},
  numpages = {7},
  year = {2021},
  month = {Jan},
  publisher = {American Physical Society},
}

@Article{Laliena_2018,
	author = {Laliena, Victor and Campo, Javier},
	journal = {Journal of Physics A: Mathematical and Theoretical},
	month = {jul},
	number = {32},
	pages = {325203},
	publisher = {IOP Publishing},
	title = {An improved discretization of {S}chr{\"o}dinger-like radial equations},
	volume = {51},
	year = {2018},
}

@book{Landau_2021,
	address = {Cambridge},
	author = {Landau, David and Binder, Kurt},
	dp = {Cambridge University Press},
	et = {5},
	isbn = {9781108490146},
	publisher = {Cambridge University Press},
	title = {A Guide to Monte Carlo Simulations in Statistical Physics},
	year = {2021},
}

@Article{Ceperley_1995,
  title = {Path integrals in the theory of condensed helium},
  author = {Ceperley, D. M.},
  journal = {Rev. Mod. Phys.},
  volume = {67},
  issue = {2},
  pages = {279--355},
  numpages = {0},
  year = {1995},
  month = {Apr},
  publisher = {American Physical Society},
}

@Article{Boninsegni_2006,
  title = {Worm algorithm and diagrammatic {M}onte {C}arlo: {A} new approach to continuous-space path integral {M}onte {C}arlo simulations},
  author = {Boninsegni, M. and Prokof'ev, N. V. and Svistunov, B. V.},
  journal = {Phys. Rev. E},
  volume = {74},
  issue = {3},
  pages = {036701},
  numpages = {16},
  year = {2006},
  month = {Sep},
  publisher = {American Physical Society},
}

@Article{Del_Maestro_2014,
  title = {Quantum {M}onte {C}arlo measurement of the chemical potential of ${}^{4}\text{He}$},
  author = {Herdman, C. M. and Rommal, A. and Del Maestro, A.},
  journal = {Phys. Rev. B},
  volume = {89},
  issue = {22},
  pages = {224502},
  numpages = {7},
  year = {2014},
  month = {Jun},
  publisher = {American Physical Society},
}

@Article{Cinti_2017,
  title = {Phases of dipolar bosons in a bilayer geometry},
  author = {Cinti, Fabio and Wang, Daw-Wei and Boninsegni, Massimo},
  journal = {Phys. Rev. A},
  volume = {95},
  issue = {2},
  pages = {023622},
  numpages = {6},
  year = {2017},
  month = {Feb},
  publisher = {American Physical Society},
}

@Article{Zimmerman_2022,
    author = {Michal Zimmerman and Ronen Rapaport and Snir Gazit},
    journal = {Proceedings of the National Academy of Sciences},
    number = {30},
    pages = {e2205845119},
    title = {Collective interlayer pairing and pair superfluidity in vertically stacked layers of dipolar excitons},
    volume = {119},
    year = {2022},
}

@article{Belgaonkar_2025,
  title = {Trimer superfluidity of antiparallel dipolar excitons in a bilayer heterostructure},
  author = {Belgaonkar, Pradyumna P. and Zimmerman, Michal and Gazit, Snir and Orgad, Dror},
  journal = {Phys. Rev. B},
  volume = {113},
  issue = {6},
  pages = {064512},
  numpages = {13},
  year = {2026},
  month = {Feb},
  publisher = {American Physical Society},
  doi = {10.1103/zn5r-2vrb},
  url = {https://link.aps.org/doi/10.1103/zn5r-2vrb}
}

@Article{Fogler_2021,
  title = {Ground and excited states of coupled exciton liquids in electron-hole quadrilayers},
  author = {Xu, Chao and Fogler, Michael M.},
  journal = {Phys. Rev. B},
  volume = {104},
  issue = {19},
  pages = {195430},
  numpages = {11},
  year = {2021},
  month = {Nov},
  publisher = {American Physical Society},
}

@Article{Rapaport_2020,
  title = {Attractive interactions, molecular complexes, and polarons in coupled dipolar exciton fluids},
  author = {Hubert, C. and Cohen, K. and Ghazaryan, A. and Lemeshko, M. and Rapaport, R. and Santos, P. V.},
  journal = {Phys. Rev. B},
  volume = {102},
  issue = {4},
  pages = {045307},
  numpages = {12},
  year = {2020},
  month = {Jul},
  publisher = {American Physical Society},
}

@Article{Hubert_2019,
  title = {Attractive Dipolar Coupling between Stacked Exciton Fluids},
  author = {Hubert, Colin and Baruchi, Yifat and Mazuz-Harpaz, Yotam and Cohen, Kobi and Biermann, Klaus and Lemeshko, Mikhail and West, Ken and Pfeiffer, Loren and Rapaport, Ronen and Santos, Paulo},
  journal = {Phys. Rev. X},
  volume = {9},
  issue = {2},
  pages = {021026},
  numpages = {12},
  year = {2019},
  month = {May},
  publisher = {American Physical Society},
}

@Article{Choksy_2021,
  title = {Attractive and repulsive dipolar interaction in bilayers of indirect excitons},
  author = {Choksy, D. J. and Xu, Chao and Fogler, M. M. and Butov, L. V. and Norman, J. and Gossard, A. C.},
  journal = {Phys. Rev. B},
  volume = {103},
  issue = {4},
  pages = {045126},
  numpages = {10},
  year = {2021},
  month = {Jan},
  publisher = {American Physical Society},
}

@Article{IBJ_2024,
	author = {Amit Jash and Michael Stern and Subhradeep Misra and Vladimir Umansky and Israel Bar Joseph},
	journal = {Science Advances},
	number = {33},
	pages = {eado8763},
	title = {Giant hyperfine interaction between a dark exciton condensate and nuclei},
	volume = {10},
	year = {2024},
}

@Article{Astrakharchik_2021,
  title = {Quantum phase transition of a two-dimensional quadrupolar system},
  author = {Astrakharchik, G. E. and Kurbakov, I. L. and Sychev, D. V. and Fedorov, A. K. and Lozovik, Yu. E.},
  journal = {Phys. Rev. B},
  volume = {103},
  issue = {14},
  pages = {L140101},
  numpages = {7},
  year = {2021},
  month = {Apr},
  publisher = {American Physical Society},
}

@Article{Deng_2024,
	author = {Eunice Paik and Long Zhang and Kin Fai Mak and Jie Shan and Hui Deng},
	journal = {Adv. Opt. Photon.},
	month = {Dec},
	number = {4},
	pages = {1064--1132},
	publisher = {Optica Publishing Group},
	title = {Excitons and polaritons in two-dimensional transition metal dichalcogenides: a tutorial},
	volume = {16},
	year = {2024},
}

@Article{Holleitner_2024,
	author = {Brotons-Gisbert, Mauro and Gerardot, Brian D. and Holleitner, Alexander W. and Wurstbauer, Ursula},
	da = {2024/09/01},
	id = {Brotons-Gisbert2024},
	isbn = {1938-1425},
	journal = {MRS Bulletin},
	number = {9},
	pages = {914--931},
	title = {Interlayer and {M}oir{\'e} excitons in atomically thin double layers: From individual quantum emitters to degenerate ensembles},
	volume = {49},
	year = {2024},
}

@Article{Gazit_2016,
  title = {Collective Modes in a Quantum Solid},
  author = {Gazit, Snir and Podolsky, Daniel and Nonne, Heloise and Auerbach, Assa and Arovas, Daniel P.},
  journal = {Phys. Rev. Lett.},
  volume = {117},
  issue = {8},
  pages = {085302},
  numpages = {5},
  year = {2016},
  month = {Aug},
  publisher = {American Physical Society},
}

@article{Podolsky_melting_2026,
  title = {Defect-mediated melting of square-lattice solids},
  author = {Grampel, William and Podolsky, Daniel},
  journal = {Phys. Rev. B},
  volume = {113},
  issue = {21},
  pages = {214109},
  numpages = {12},
  year = {2026},
  month = {Jun},
  publisher = {American Physical Society},
}

@Article{Lambrecht2012biMoSe2,
  title = {Quasiparticle band structure calculation of monolayer, bilayer, and bulk {M}o{S}$_{2}$},
  author = {Cheiwchanchamnangij, Tawinan and Lambrecht, Walter R. L.},
  journal = {Phys. Rev. B},
  volume = {85},
  issue = {20},
  pages = {205302},
  numpages = {4},
  year = {2012},
  month = {May},
  publisher = {American Physical Society},
}

@Article{Berkelbach2013mono,
  title = {Theory of neutral and charged excitons in monolayer transition metal dichalcogenides},
  author = {Berkelbach, Timothy C. and Hybertsen, Mark S. and Reichman, David R.},
  journal = {Phys. Rev. B},
  volume = {88},
  issue = {4},
  pages = {045318},
  numpages = {6},
  year = {2013},
  month = {Jul},
  publisher = {American Physical Society},
}

@Article{Glazov_2019,
  title = {Colloquium: Excitons in atomically thin transition metal dichalcogenides},
  author = {Wang, Gang and Chernikov, Alexey and Glazov, Mikhail M. and Heinz, Tony F. and Marie, Xavier and Amand, Thierry and Urbaszek, Bernhard},
  journal = {Rev. Mod. Phys.},
  volume = {90},
  issue = {2},
  pages = {021001},
  numpages = {25},
  year = {2018},
  month = {Apr},
  publisher = {American Physical Society},
}

@Article{Chomaz_2023,
	author = {Chomaz, Lauriane and Ferrier-Barbut, Igor and Ferlaino, Francesca and Laburthe-Tolra, Bruno and Lev, Benjamin L and Pfau, Tilman},
	journal = {Reports on Progress in Physics},
	month = {dec},
	number = {2},
	pages = {026401},
	publisher = {IOP Publishing},
	title = {Dipolar physics: a review of experiments with magnetic quantum gases},
	volume = {86},
	year = {2022},
}

@Article{Baranov_2012,
	author = {Baranov, M. A. and Dalmonte, M. and Pupillo, G. and Zoller, P.},
	da = {2012/09/12},
	date = {2012/09/12},
	isbn = {0009-2665},
	journal = {Chemical Reviews},
	month = {09},
	number = {9},
	pages = {5012--5061},
	publisher = {American Chemical Society},
	title = {Condensed Matter Theory of Dipolar Quantum Gases},
	volume = {112},
	year = {2012},
	year1 = {2012},
}

@Article{Wang_2019,
    author = {Jue Wang and Jenny Ardelean and Yusong Bai and Alexander Steinhoff and Matthias Florian and Frank Jahnke and Xiaodong Xu and Mackillo Kira and James Hone and X.-Y. Zhu},
    journal = {Science Advances},
    number = {9},
    pages = {eaax0145},
    title = {Optical generation of high carrier densities in 2{D} semiconductor heterobilayers},
    volume = {5},
    year = {2019},
}

@Article{Bishwajit_2017,
  title = {Exciton condensate in bilayer transition metal dichalcogenides: Strong coupling regime},
  author = {Debnath, Bishwajit and Barlas, Yafis and Wickramaratne, Darshana and Neupane, Mahesh R. and Lake, Roger K.},
  journal = {Phys. Rev. B},
  volume = {96},
  issue = {17},
  pages = {174504},
  numpages = {8},
  year = {2017},
  month = {Nov},
  publisher = {American Physical Society},
}

@Article{MacDonald_2018,
  title = {Theory of optical absorption by interlayer excitons in transition metal dichalcogenide heterobilayers},
  author = {Wu, Fengcheng and Lovorn, Timothy and MacDonald, A. H.},
  journal = {Phys. Rev. B},
  volume = {97},
  issue = {3},
  pages = {035306},
  numpages = {10},
  year = {2018},
  month = {Jan},
  publisher = {American Physical Society},
}

@Article{Goryca_2019,
	author = {Goryca, M. and Li, J. and Stier, A. V. and Taniguchi, T. and Watanabe, K. and Courtade, E. and Shree, S. and Robert, C. and Urbaszek, B. and Marie, X. and Crooker, S. A.},
	id = {Goryca2019},
	isbn = {2041-1723},
	journal = {Nature Communications},
	number = {1},
	pages = {4172},
	title = {Revealing exciton masses and dielectric properties of monolayer semiconductors with high magnetic fields},
	volume = {10},
	year = {2019},
}

@Article{Heinz_2022,
	author = {Karni, Ouri and Barr{\'e}, Elyse and Pareek, Vivek and Georgaras, Johnathan D. and Man, Michael K. L. and Sahoo, Chakradhar and Bacon, David R. and Zhu, Xing and Ribeiro, Henrique B. and O'Beirne, Aidan L. and Hu, Jenny and Al-Mahboob, Abdullah and Abdelrasoul, Mohamed M. M. and Chan, Nicholas S. and Karmakar, Arka and Winchester, Andrew J. and Kim, Bumho and Watanabe, Kenji and Taniguchi, Takashi and Barmak, Katayun and Mad{\'e}o, Julien and da Jornada, Felipe H. and Heinz, Tony F. and Dani, Keshav M.},
	isbn = {1476-4687},
	journal = {Nature},
	number = {7900},
	pages = {247--252},
	title = {Structure of the moir{\'e} exciton captured by imaging its electron and hole},
	volume = {603},
	year = {2022},
}

@article{Knap2025Diffusion,
	author = {Upadhyay, Pranshoo and Su{\'a}rez-Forero, Daniel G. and Huang, Tsung-Sheng and Mehrabad, Mahmoud Jalali and Gao, Beini and Sarkar, Supratik and Session, Deric and Watanabe, Kenji and Taniguchi, Takashi and Zhou, You and Knap, Michael and Hafezi, Mohammad},
	doi = {10.1126/science.ads5266},
	journal = {Science},
	number = {6783},
	pages = {394--398},
	publisher = {American Association for the Advancement of Science},
	title = {Giant enhancement of exciton diffusion near an electronic Mott insulator},
	ty = {JOUR},
	url = {https://doi.org/10.1126/science.ads5266},
	volume = {391},
	year = {2026},
    }

@misc{Knap_2024,
      title={Gate-tunable {B}ose-{F}ermi mixture in a strongly correlated moir\'e bilayer electron system}, 
      author={Amine Ben Mhenni and Wilhelm Kadow and Mikołaj J. Metelski and Adrian O. Paulus and Alain Dijkstra and Kenji Watanabe and Takashi Taniguchi and Seth Ariel Tongay and Matteo Barbone and Jonathan J. Finley and Michael Knap and Nathan P. Wilson},
      eprint={2410.07308},
      archivePrefix={arXiv},
    year={2024},
}

@article{Zerba_2024,
  title = {Realizing Topological Superconductivity in Tunable {B}ose-{F}ermi Mixtures with Transition Metal Dichalcogenide Heterostructures},
  author = {Zerba, Caterina and Kuhlenkamp, Clemens and Imamo{\u g}lu, Ata{\c c} and Knap, Michael},
  journal = {Phys. Rev. Lett.},
  volume = {133},
  issue = {5},
  pages = {056902},
  numpages = {8},
  year = {2024},
  publisher = {American Physical Society},
}

@article{Zerba_2025,
  title = {Tuning Transport in Solid-State {B}ose-{F}ermi Mixtures by {F}eshbach Resonances},
  author = {Zerba, Caterina and Kuhlenkamp, Clemens and Mangeolle, L\'eo and Knap, Michael},
  journal = {Phys. Rev. Lett.},
  volume = {134},
  issue = {12},
  pages = {126502},
  numpages = {7},
  year = {2025},
  publisher = {American Physical Society},
}

@Article{Mak2021EI,
	author = {Ma, Liguo and Nguyen, Phuong X. and Wang, Zefang and Zeng, Yongxin and Watanabe, Kenji and Taniguchi, Takashi and MacDonald, Allan H. and Mak, Kin Fai and Shan, Jie},
	id = {Ma2021},
	isbn = {1476-4687},
	journal = {Nature},
	number = {7882},
	pages = {585--589},
	title = {Strongly correlated excitonic insulator in atomic double layers},
	volume = {598},
	year = {2021},

}

@Article{Mak2022EI,
	author = {Gu, Jie and Ma, Liguo and Liu, Song and Watanabe, Kenji and Taniguchi, Takashi and Hone, James C. and Shan, Jie and Mak, Kin Fai},
	isbn = {1745-2481},
	journal = {Nature Physics},
	number = {4},
	pages = {395--400},
	title = {Dipolar excitonic insulator in a moir{\'e} lattice},
	volume = {18},
	year = {2022},
}

@Article{Watanabe2022EI,
	author = {Zhang, Zuocheng and Regan, Emma C. and Wang, Danqing and Zhao, Wenyu and Wang, Shaoxin and Sayyad, Mohammed and Yumigeta, Kentaro and Watanabe, Kenji and Taniguchi, Takashi and Tongay, Sefaattin and Crommie, Michael and Zettl, Alex and Zaletel, Michael P. and Wang, Feng},
	id = {Zhang2022},
	isbn = {1745-2481},
	journal = {Nature Physics},
	number = {10},
	pages = {1214--1220},
	title = {Correlated interlayer exciton insulator in heterostructures of monolayer {W}{S}e$_2$ and moir{\'e} {W}{S}$_2$/{W}{S}e$_2$},
	volume = {18},
	year = {2022},
}

@Article{Watanabe2020Wigner,
	author = {Regan, Emma C. and Wang, Danqing and Jin, Chenhao and Bakti Utama, M. Iqbal and Gao, Beini and Wei, Xin and Zhao, Sihan and Zhao, Wenyu and Zhang, Zuocheng and Yumigeta, Kentaro and Blei, Mark and Carlstr{\"o}m, Johan D. and Watanabe, Kenji and Taniguchi, Takashi and Tongay, Sefaattin and Crommie, Michael and Zettl, Alex and Wang, Feng},
	isbn = {1476-4687},
	journal = {Nature},
	number = {7799},
	pages = {359--363},
	title = {Mott and generalized {W}igner crystal states in {W}{S}e$_2$/{W}{S}$_2$ moir{\'e} superlattices},
	volume = {579},
	year = {2020},
}

@Article{Watanabe2020Mott,
	author = {Tang, Yanhao and Li, Lizhong and Li, Tingxin and Xu, Yang and Liu, Song and Barmak, Katayun and Watanabe, Kenji and Taniguchi, Takashi and MacDonald, Allan H. and Shan, Jie and Mak, Kin Fai},
	isbn = {1476-4687},
	journal = {Nature},
	number = {7799},
	pages = {353--358},
	title = {Simulation of {H}ubbard model physics in {W}{S}e$_2$/{W}{S}$_2$ moir{\'e} superlattices},
	volume = {579},
	year = {2020},
}

@Article{Mak2020Mott,
	author = {Xu, Yang and Liu, Song and Rhodes, Daniel A. and Watanabe, Kenji and Taniguchi, Takashi and Hone, James and Elser, Veit and Mak, Kin Fai and Shan, Jie},
	isbn = {1476-4687},
	journal = {Nature},
	number = {7833},
	pages = {214--218},
	title = {Correlated insulating states at fractional fillings of moir{\'e} superlattices},
	volume = {587},
	year = {2020},
}

@Article{Imamoglu2020Mott,
	author = {Shimazaki, Yuya and Schwartz, Ido and Watanabe, Kenji and Taniguchi, Takashi and Kroner, Martin and Imamo{\u g}lu, Ata{\c c}},
	id = {Shimazaki2020},
	isbn = {1476-4687},
	journal = {Nature},
	number = {7804},
	pages = {472--477},
	title = {Strongly correlated electrons and hybrid excitons in a moir{\'e} heterostructure},
	volume = {580},
	year = {2020},
}

@Article{Wang2020Mott,
	author = {Wang, Lei and Shih, En-Min and Ghiotto, Augusto and Xian, Lede and Rhodes, Daniel A. and Tan, Cheng and Claassen, Martin and Kennes, Dante M. and Bai, Yusong and Kim, Bumho and Watanabe, Kenji and Taniguchi, Takashi and Zhu, Xiaoyang and Hone, James and Rubio, Angel and Pasupathy, Abhay N. and Dean, Cory R.},
	isbn = {1476-4660},
	journal = {Nature Materials},
	number = {8},
	pages = {861--866},
	title = {Correlated electronic phases in twisted bilayer transition metal dichalcogenides},
	volume = {19},
	year = {2020},
}

@Article{Imamoglu2021Feshbach,
	author = {Ido Schwartz and Yuya Shimazaki and Clemens Kuhlenkamp and Kenji Watanabe and Takashi Taniguchi and Martin Kroner and Ata{\c c} Imamo{\u g}lu},
	journal = {Science},
	number = {6565},
	pages = {336-340},
	title = {Electrically tunable {F}eshbach resonances in twisted bilayer semiconductors},
    volume = {374},
	year = {2021},
}

@Article{Imamoglu2022Feshbach,
  title = {Tunable {F}eshbach Resonances and Their Spectral Signatures in Bilayer Semiconductors},
  author = {Kuhlenkamp, Clemens and Knap, Michael and Wagner, Marcel and Schmidt, Richard and Imamo{\u g}lu, Ata{\c c}},
  journal = {Phys. Rev. Lett.},
  volume = {129},
  issue = {3},
  pages = {037401},
  numpages = {7},
  year = {2022},
  publisher = {American Physical Society},
}

@article{Kang_2013,
    author = {Kang, Jun and Tongay, Sefaattin and Zhou, Jian and Li, Jingbo and Wu, Junqiao},
    title = {Band offsets and heterostructures of two-dimensional semiconductors},
    journal = {Applied Physics Letters},
    volume = {102},
    pages = {012111},
    year = {2013},
}

@software{trilayer_zenodo,
  author       = {Zimmerman, Michal and Podolsky, Daniel and Rapaport, Ronen and Gazit, Snir},
  title        = {Quadrupolar and dipolar phases of excitons in transition-metal dichalcogenide trilayer heterostructures - data [Dataset]},
  year         = 2025,
  publisher    = {Zenodo},
  doi          ={https://doi.org/10.5281/zenodo.18075094.},
}

\end{document}